\documentclass[aps,prb,amsmath,amssymb,superscriptaddress,reprint,showpacs]{revtex4-1}
\makeatother
\usepackage{graphicx}% Include figure files
\usepackage{dcolumn}% Align table columns on decimal point
\usepackage{bm}% bold math
\usepackage{float}
\usepackage{sidecap}
\usepackage{color} 
\usepackage{relsize}
\newcounter{eqn}

\newcommand{\putindeepbox}[2][0.7\baselineskip]{{%
    \setbox0=\hbox{#2}%
    \setbox0=\vbox{\noindent\hsize=\wd0\unhbox0}
    \@tempdima=\dp0
    \advance\@tempdima by \ht0
    \advance\@tempdima by -#1\relax
    \dp0=\@tempdima
    \ht0=#1\relax
    \box0
}}

\newcommand{\bfr}{ {\bf r}} % definitions of r as a vector
\newcommand{\bfrp}{ {\bf r'}} % definitions of r' as a vector
\newcommand{\bfR}{ {\bf R}} % definitions of R as a vector
\newcommand{\bfk}{ {\bf k}} % definitions of k as a vector
\mathchardef\mhyphen="2D

\makeatother
\begin{document}

\preprint{AIP/123-QED}

\title[Self-Consistent Green's Function Embedding for Advanced Electronic Structure Methods Based on DMFT]
      {Self-Consistent Green's Function Embedding for Advanced Electronic Structure Methods Based on a Dynamical Mean-Field Concept}% Force line breaks with \\
% \thanks{Footnote to title of article.}

\author{Wael Chibani}
%  \altaffiliation[Also at ]{Physics Department, XYZ University.}%Lines break automatically or can be forced with \\
 \email{chibani@fhi-berlin.mpg.de}
 \affiliation{ 
Fritz-Haber-Institut der Max-Planck-Gesellschaft, Faradayweg 4-6, D-14195 Berlin, Germany%\\This line break forced with \textbackslash\textbackslash
}
\author{Xinguo Ren}%
%  \email{Second.Author@institution.edu.}
 \affiliation{ 
Fritz-Haber-Institut der Max-Planck-Gesellschaft, Faradayweg 4-6, D-14195 Berlin, Germany%\\This line break forced with \textbackslash\textbackslash
}
\affiliation{%
Key Laboratory of Quantum Information, University of Science and Technology of China, Hefei 230026, China
}%

\author{Matthias Scheffler}
%  \homepage{http://www.Second.institution.edu/~Charlie.Author.}
 \affiliation{ 
Fritz-Haber-Institut der Max-Planck-Gesellschaft, Faradayweg 4-6, D-14195 Berlin, Germany%\\This line break forced with \textbackslash\textbackslash
}

\author{Patrick Rinke}
%  \homepage{http://www.Second.institution.edu/~Charlie.Author.}
 \affiliation{ 
Fritz-Haber-Institut der Max-Planck-Gesellschaft, Faradayweg 4-6, D-14195 Berlin, Germany%\\This line break forced with \textbackslash\textbackslash
}
\affiliation{COMP/Department of Applied Physics, Aalto University, P.O. Box 11100, Aalto FI-00076, Finland}

\date{\today}% It is always \today, today,
             %  but any date may be explicitly specified

\begin{abstract}

We present an embedding scheme for periodic systems that facilitates 
the treatment of the physically important part  (here the unit cell) 
with advanced electronic-structure methods, that are computationally too expensive 
for periodic systems. The rest of the periodic system is treated with 
computationally less demanding approaches, e.g., Kohn-Sham density-functional theory, in a self-consistent manner. 
Our scheme is based on the concept of dynamical mean-field theory  (DMFT) formulated in terms of Green's functions. 
In contrast to the original DMFT formulation for correlated model Hamiltonians, 
we here consider the unit cell as local embedded cluster in a first-principles way, 
that includes all electronic degrees of freedom. Our real-space dynamical mean-field embedding (RDMFE) scheme features two nested Dyson equations, one for the embedded cluster 
and another for the periodic surrounding. The total energy is computed from the resulting Green's functions.
The performance of our scheme is demonstrated by treating the embedded region 
with hybrid functionals and many-body perturbation theory in the \textit{GW} approach for simple bulk systems. 
The total energy and the density of states converge rapidly with respect to 
the computational parameters and approach their bulk limit with increasing 
cluster (i.e., unit cell) size.

% Valid PACS numbers may be entered using the \verb+\pacs{#1}+ command.
\end{abstract}

\pacs{Valid PACS appear here}% PACS, the Physics and Astronomy
                             % Classification Scheme.
%Use showkeys class option if keyword
                              %display desired
\maketitle

\section{Introduction}

Density-functional theory (DFT) has become a widely applied electronic-structure theory method due to the balance between accuracy and 
computational efficiency of local and semi local approximations such as the local-density (LDA) and generalized gradient approximations (GGA). 
However, LDA and GGAs suffer from certain intrinsic limitations such as the self-interaction error \cite{Perdew/Zunger:1981,Perdew/etal:1982,Mori_Sanchez/etal2006}, 
the absence of the derivative discontinuity in the exchange-correlation potential \cite{Perdew/Levy:1983,Sham/Schlueter:1983,Mori-Sanchez_1/etal:2009}, 
the lack of long-range van der Waals interactions \cite{Gunnarsson/Lundqvist:1976,Dobson/Wang:1999,Tkatchenko/Scheffler:2009}, and the absence of image effects 
\cite{White/Godby/Rieger/Needs:1997,Thygesen/Rubio:2009,Freysoldt/Rinke/Scheffler:2009}. 
These shortcomings limit the predictive power of LDA and GGAs, in particular for localised electrons as found in $d$- or $f$-electron systems 
\cite{Leung/etal:1988,Zaanen/etal:1988,Pickett:1989, Mattheiss_1:1972,Mattheiss_2:1972} or for adsorbates and surfaces. 
\cite{Feibelman:2001,MehdaouiandKluener:2007,Mehdaoui/etal:2007,MehdaouiandKluener:2008} 
Furthermore, DFT is inherently a ground-state method and therefore of limited applicability for excited states and spectra. 
More advanced electronic-structure methods that overcome one or several of the mentioned shortcomings exist, but they are typically computationally 
much more demanding and thus limited to small systems sizes or a subset of electronic degrees of freedom. To overcome the \emph{efficiency-accuracy} conundrum, 
much effort has been devoted to combine the best of both worlds, that is to merge local and semilocal DFT approximations (DFA) with advanced electronic 
methods\cite{Zgid/Chan:2011,Singh/Kollman:1986,Field/Bash/Martin:1990,Maseras/Morokuma:1995,Scheffler/etal:1985,Scheffler/Bormet:1994,Whitten/Pakkanen:1980,Huang/Carter:2011,Hu/Reuter/Scheffler:2007,Ren/Rinke/Scheffler:2009,Metzner/Vollhardt:1989,Georges/Kotliar:1992,Georges/etal:1996,Knizia/Chan:2012}.

\begin{figure}[htp]
    \centering
    \includegraphics[scale=0.25]{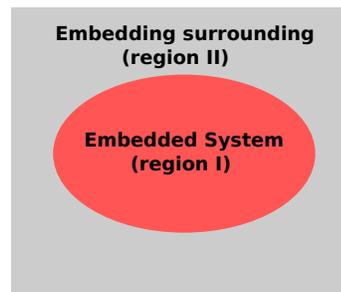}
    \caption{Schematic depiction of the embedding concept. The embedded region (red) is treated with a
             more accurate theory,  whereas the surrounding (grey) is calculated with less accurate 
             and thus computationally less expensive theories. The key challenge of our and all previous embedding approaches is 
             the appropriate treatment of the red/grey boundary.}
    \label{fig:embedding_cartoon}
\end{figure}

We here advocate the concept of embedding. In this divide and conquer approach, the full system is divided into two parts: a small embedded region, which is treated with advanced, 
computationally demanding approaches, and an embedding environment that is treated with computationally more efficient approaches.  A schematic illustration of
the embedding concept is shown in Fig.~\ref{fig:embedding_cartoon}.
%

%marker
Following this general principle, various embedding schemes have been developed in the past
\cite{Zgid/Chan:2011,Singh/Kollman:1986,Field/Bash/Martin:1990,Maseras/Morokuma:1995,Scheffler/etal:1985,Scheffler/Bormet:1994,Whitten/Pakkanen:1980,Huang/Carter:2011,Hu/Reuter/Scheffler:2007,Ren/Rinke/Scheffler:2009,Metzner/Vollhardt:1989,Georges/Kotliar:1992,Georges/etal:1996,Knizia/Chan:2012,Berger/etal:2014}. 
They differ in scope (i.e., area of application), on how they treat the coupling between the embedded region and the surrounding, 
and in the approaches used to describe the two regions. In (bio)chemistry, for example, one of the most popular embedding schemes 
combines quantum mechanics (QM) and classical molecular mechanics (MM). The embedded region is treated quantum mechanically and the surrounding 
by MM. \cite{Singh/Kollman:1986,Field/Bash/Martin:1990,Maseras/Morokuma:1995,Scheffler/etal:1985,Scheffler/Bormet:1994} 
In surface science, fully quantum mechanical schemes are more prevalent, e.g., for the description of surface adsorbates. 
They divide space into regions for advanced and less advanced electronic-structure approaches and differ mostly on how these two regions are coupled, 
e.g. through maximal exchange overlap \cite{Whitten/Pakkanen:1980}, density embedding \cite{Huang/Carter:2011} or cluster extrapolation \cite{Hu/Reuter/Scheffler:2007,Ren/Rinke/Scheffler:2009}. In solid-state physics, dynamical mean-field theory \cite{Metzner/Vollhardt:1989,Georges/Kotliar:1992,Georges/etal:1996} offers a natural embedding framework by mapping an infinite,
correlated lattice model into an impurity model \cite{Anderson:1961} immersed into a self-consistently determined mean-field bath. 
When DFA is chosen as the mean field, DMFT becomes material specific\cite{Georges/etal:1996,Kotliar/etal:2006,Held:2007}.
The embedding is achieved by means of Green's functions facilitating the calculation of spectra, band structures, but also phase diagrams.
Recently, Zgid and Chan \cite{Zgid/Chan:2011} 
proposed to use DMFT as an embedding framework for quantum-chemistry approaches such as the configuration-interaction (CI) method. 
They since proposed a simplified DMFT scheme based on density-matrix embedding to access static properties (e.g., the ground-state energy and its derivatives)\cite{Knizia/Chan:2012}.
However, at present all DMFT approaches use a down-folding procedure to a low energy subspace, which is treated on the model-Hamiltonian level.

We here extend the DMFT concept to couple two \emph{first-principles} regions. Our idea is similar to that of Zgid and Chan\cite{Zgid/Chan:2011}, 
but we explore the possibility of using DMFT as a general embedding scheme for advanced first-principles electronic-structure methods. These can be advanced DFT exchange-correlation functionals or excited-state methods based on the $GW$ approach 
\cite{Hedin:1965}, which are still computationally too expensive for large-scale systems. The difference to previous DMFT schemes is that we treat the unit cell as the local, embedded region, that is coupled to the rest of the periodic system via the DMFT framework. The advantage of our approach is that all electrons in the embedded region are treated on the same quantum mechanical level, which removes the arbitrariness in the definition of the down-folded subspace and does not require any double counting corrections. Furthermore, our approach includes non-local interactions in the embedded region, whose size can be systematically converged to the thermodynamic limit. 

We here present the concept of our real-space dynamical mean-field embedding (RDMFE) approach and its implementation in the all-electron Fritz Haber Institute \textit{ab initio} molecular simulations (FHI-aims) code \cite{Blum/etal:2009,Havu/etal:2009,Ren/etal:2012}. We first benchmark it for hybrid density functionals for which we have a periodic reference \cite{Levchenko/etal:2015}. Then we apply our scheme to the $GW$ approach. Due to its intrinsic self-consistency, our RDMFE approach yields a self-consistent $GW$ solution, which is a much coveted approach for solids right now. 
While fully self-consistent $GW$ implementations 
for molecules are slowly emerging \cite{Stan/Dahlen/Leeuwen:2006,Rostgaard/Jacobsen/Thygesen:2010,Caruso/etal:2012,Caruso/etal:2013_tech,Caruso/etal:2013_H2,Koval/etal:2014}, 
we are only aware of one recent implementation of 
fully self-consistent $GW$ for solids \cite{Kutepov:2009,Kutepov/etal:2012}, which is, however, limited to small unit cells due to its computational expense. An alternative, approximate way to achieve self-consistency within $GW$ is the so-called quasiparticle self-consistent $GW$ (QPscGW) scheme, \cite{Schilfgaarde/Kotani/Faleev:2006,Kotani/etal:2007} 
which was recently been widely applied to solids. We here present self-consistent $GW$ spectra and ground-state energies for simple solids.

The rest of the paper is organized as follows: The detailed formalism of our Green's-function based embedding scheme is derived in Section~\ref{sec:method}. Section~\ref{sec:results} presents benchmark results including both total energies and band structures for simple bulk systems. In Section~\ref{sec:discussion} we contrast our approach with the aforementioned embedding schemes. Section~\ref{sec:summary} concludes the paper.

\section{\label{sec:method}  Self-Consistent Green's function Embedding in Real Space}

\subsection{General concept}
In its original formulation, \cite{Georges/etal:1996} DMFT is a Green's function method for correlated model 
Hamiltonians that uses the locality of the electronic interaction
to embed a local on-site region of the Hubbard lattice -- typically a 
 single $d$- or $f$-level -- into a periodic electronic bath defining a 
 self-consistent scheme. Treating the on-site region
 locally facilitates the use of computationally very demanding and at the same time very accurate methods 
 such as continuous time quantum Monte-Caro \cite{Werner/etal:2006}, direct diagonalization or 
 renormalization group techniques \cite{Zgid/etal:2012}. The  localized region is then coupled through a hybridization self-energy
 to the surrounding electronic bath, which is treated with computationally more efficient methods.
 In the last few years, DMFT has proven to be very successful in describing the 
spectral properties of solids with localised $d$- and $f$-states \cite{Savrasov/etal:2001,Held/etal:2001,Amadon/etal:2006,Tomczak/etal:2008}.
 However, present versions of DMFT also suffer from certain shortcomings. For example, a double counting problem
 arises when local or semilocal DFA is used to calculate the electronic surrounding, because the DFA contribution of the localised manifold, which would have to be subtracted in order to not count it twice, is not known \cite{Karolak:2010}. Different double counting schemes can give vastly different answers in DFA+DMFT \cite{Lichtenstein:1998,Anisimov/etal:1997}. 
%  Also, the calculation of the hybridization self-energy is a delicate task \cite{Liebsch/etal:2012}. Some high level methods for the on-site region require fitting to the bath degrees of freedom \PR{I didn't really understand this point. Can you check that my version of this is actually correct.}. 
% To make the fitting error as small as possible, the number of bath fitting levels can be very large making the calculation computationally costly again. 
Last, but not least, the choice of the local manifold can have a considerable impact on the result, in particular when the localised states hybridise with delocalised states in the system \cite{Vaugier/etal:2012,Nilsson/etal:2013}.
In principle, one could use renormalisation groups to derive the most suitable low energy Hamiltonian for the localised region systematically. However, in practice this is not tractable computationally.
 
\begin{figure*}[htp]
	    \centering
		\includegraphics[scale=0.25]{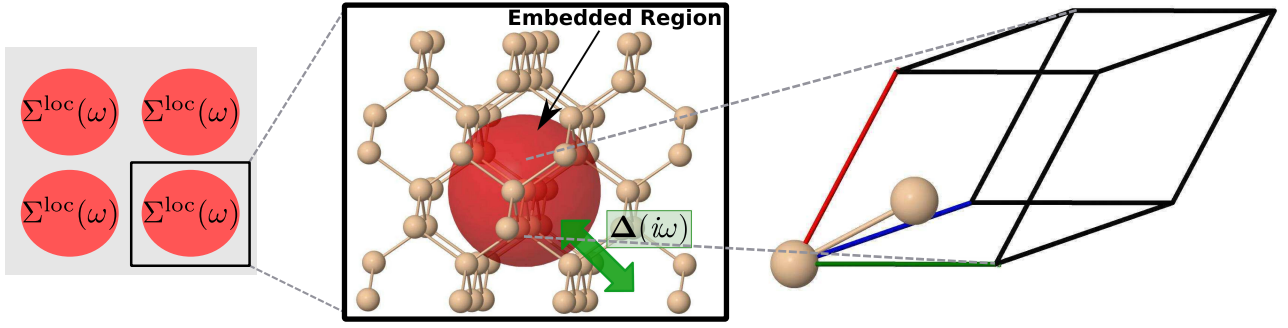}
		\caption{The DMFT embedding concept for a Si unit cell. The atoms in the unit cell (red region) constitute the embedded sub-manifold. 
		Each unit cell of the periodic system is treated as a localized region, i.e. only local interactions $\Sigma^{\rm loc}$
		are treated. The unit cells are coupled to the rest of the system via the hybridization self-energy $\Delta(i\omega)$ (green arrow).  
% 		\PR{I see that you went back to an earlier version of this figure. This makes sense, because the previous version was confusing. But now you don't include the unit cell anymore. Can you add a third panel again on the right that is a blow-up of the red region in the middle panel? In the third panel you should just show the primitive unit cell of silicon (i.e., the parallelepiped that is the unit cell with two atoms inside)}
% 		\Wael{I still have to do this!}
% \PR{I really like the right-hand-side of this figure now! However, now the left-hand-side is way to small! Can you either enlarge the left panel or reduce the number of read ellipsoids to four (and thus make everything larger)?} \PR{To make the figure even clearer, you could/should include a 2nd zoom level. I would replace the red sphere in the right panel by the unit cell. Draw the unit cell edges, but keep the transparent red shading. Then you do the 2nd zoom-out and show only the silicon unit cell with two atoms inside. Then it becomes really clear that in our high-level calculations we only have a bare two atom cluster plus the additional potential $\Delta(\omega)$. If you don't have enough space in a single column, you can make this a column spanning figure by using the figure* environment.}
		  }\label{fig:dmft_concept}
	\end{figure*}

In this work, we use the concept of DMFT and formulate it as a Green's function embedding scheme for the first-principles 
% \PR{we shouldn't use both ``first-principles'' and ``first-principles'' in the same paper. Choose one or the other. If you choose ``first-principles'' you need to define it the first time you use it. } 
Hamiltonian. In contrast to the original DMFT formulation for correlated model Hamiltonians, we consider the unit cell (or any computational supercell that can span the whole space when periodically repeated) as embedded cluster in an first-principles way (see Fig.~\ref{fig:dmft_concept}). Our approach therefore  includes all electronic degrees of freedom and does not require any downfolding. Our approach also permits the charge flow from one region to the other and therefore naturally incorporates the boundary between the two regions. No special treatment is necessary for atoms on the boundary, nor is it a problem with the boundary cuts covalent bonds.

% 	\begin{figure}[htp]
% 	    \centering
% 		\includegraphics[scale=0.15]{figures/embedding/dmft_embedded_Si.eps}
% 		\caption{The atoms in the unit cell (red region) are coupled to the rest of the periodic system via the hybridization self-energy $\Delta(i\omega)$, 
% 		while only local interactions are considered}\label{fig:embedding_concept}
% 	\end{figure}
% \XR{Fig.~2 is not referred to right now.}

\subsection{Embedding scheme based on DMFT}

\subsubsection{Green's function in a non-orthogonal basis set}
The embedding framework of DMFT is most conveniently formulated in terms of Green's functions.  
In a finite (and generally nonorthogonal) basis set $\{\phi_i\}$, the Green's function
$G(\bfr, \bfrp, i\omega)$ can be expanded as,
  \begin{equation}
       G(\bfr, \bfrp, i\omega) = \sum_{i,j} \phi_i(\bfr) G_{ij}(i\omega) \phi_j(\bfrp)\, ,
%        G_{ij}(i\omega)=\int d\textbf{r}d \textbf{r}' \phi^{*}_i(\bfr) G(\bfr, \bfrp, i\omega) \phi_j (\bfrp)
  \end{equation}
where $G_{ij}(i\omega)$ is the matrix form of the Green's function.
Here we use the Green's function on the imaginary frequency axis for computational convenience and without loss of generality. For a non-interacting Hamiltonian 
$H^0_{ij}=\langle \phi_i |{\hat H}^0 |\phi_j \rangle $ (e.g., the Kohn-Sham Hamiltonian), the corresponding non-interacting Green's function 
$G^0$ in its matrix form satisfies
  \begin{equation}
     \sum_k((i\omega + \mu)S_{ik} - H^0_{ik})G^0_{kj} = \delta_{ij}\, ,
  \end{equation}
where $S_{ij}=\langle \phi_i | \phi_j \rangle$ is the overlap matrix of the basis functions, and $\mu$ is
the chemical potential. Using the Dyson equation that connects the non-interacting Green's function $G^0$ with the fully interacting one ($G$),  we obtain
  \begin{equation}
     \sum_k((i\omega + \mu)S_{ik} - H^0_{ik} -\Sigma_{ik}(i\omega))G_{kj}(i\omega) = \delta_{ij}\, ,
     \label{Eq:Green_nonortho}
  \end{equation}
where $\Sigma(i\omega)$ is the electronic self-energy.
For periodic systems, the Hamiltonian and the Green's functions are
characterized by a Bloch $\bfk$-vector in the first Brillouin zone of reciprocal space. 
Equation~\ref{Eq:Green_nonortho} thus becomes
  \begin{equation}
     \sum_k((i\omega + \mu)S_{ik}(\bfk) - H^0_{ik}(\bfk) -\Sigma_{ik}(\bfk,i\omega))G^{\rm lat}_{kj}(\bfk,i\omega) = \delta_{ij}\, ,
     \label{Eq:Green_nonortho_k}
  \end{equation}
with the lattice Green function $G^{\rm lat}(\bfk,i\omega)$.
Our implementation is based on the all-electron FHI-aims code package, \cite{Blum/etal:2009} which uses
numerical atom-centered orbitals (NAOs) as its basic functions. The basis functions $\{\phi_i\}$ will thus be
NAOs in our work.

\subsubsection{The ``on-site" Green's function for a periodic system}
The $\bfk$-dependent Green's function and self-energy in Eq.~(\ref{Eq:Green_nonortho_k}) can be Fourier-transformed
to real-space,
   \begin{align}
      G^{\rm lat}_{ij}(\bfR_i-\bfR_j,i\omega) = \frac{1}{\rm N_{1.BZ}}\sum_{\bfk}^{\rm 1.BZ}e^{i(\bfR_i-\bfR_j)\cdot \bfk} G^{\rm lat}_{ij}(\bfk,i\omega) \nonumber \\
      \Sigma_{ij}(\bfR_i-\bfR_j,i\omega) = \frac{1}{\rm N_{1.BZ}}\sum_{\bfk}^{\rm 1.BZ}e^{i(\bfR_i-\bfR_j)\cdot \bfk} \Sigma_{ij}(\bfk,i\omega) \nonumber \\
   \end{align}
where $\bfR_i$ and $\bfR_j$ are Bravais lattice vectors denoting the unit cells in which the basis functions $i$ and $j$ are
located. $N_{1\text{BZ}}$ is the number of $\bfk$-points in the first Brillouin zone (1.BZ). 
The concept of DMFT is based on the fact that the lattice self-energy becomes local, or $\bfk$-independent, 
in infinite dimension ($D=\infty$). \cite{Metzner/Vollhardt:1989} For a crystal with translational symmetry 
this implies
   \begin{equation}
       \Sigma_{ij}(\bfR_i-\bfR_j,i\omega)=\Sigma_{ij}^\text{loc}(i\omega)\delta_{\bfR_i,\bfR_j}\, .
   \end{equation}
Thus the self-energy is non-zero only if the two basis functions originate from the same unit cell. 
We call this the local (loc) or ``on-site" self-energy, following the terminology of the model-Hamiltonian studies.
In this limit,  the whole periodic system can be mapped onto an effective impurity model
of a local unit cell dynamically coupled to an effective ``external" potential arising
from the rest of the crystal.  

The first step to establish this mapping is to define the ``on-site" Green's function, i.e.,
$G_{ij}(\bfR_i-\bfR_j,i\omega)$ with $\bfR_i$=$\bfR_j$.
Using the locality of the self-energy and Eq.~(\ref{Eq:Green_nonortho_k}), we obtain the following expression 
for the onsite Green's function,
	\begin{align}
           &	G^{\mathrm{on \mhyphen site}}_{ij}(i\omega) =\frac{1}{N_{\rm 1.BZ}}\sum \limits ^{\rm 1.BZ}_{\bf k} G^{\rm lat}_{ij}(\bfk,i\omega) = \nonumber \\
           &   \frac{1}{N_{\rm 1.BZ}}\sum \limits ^{\rm 1.BZ}_{\bf k} \left[
            (i\omega + \mu)S(\bfk) - H^0(\bfk) - \Sigma^{\rm loc}(i\omega) \right]^{-1} .
%            & \frac{1}{N_{\rm 1BZ}}\sum \limits ^{\rm 1BZ}_{\bf k} G^{\rm lat}_{ij}(\bfk,i\omega)
	  \label{eq:on_site_gf}
	  \end{align}
In the DMFT context this equation is also known as the $\bfk$-integrated Dyson equation. 
So far we have not specified $H^0$. In our embedding scheme, the environment is treated by KS-DFA in the LDA or the Perdew-Burke-Ernzerhof (PBE)\cite{Perdew/Burke/Ernzerhof:1996} GGA. 
% \PR{We might have to spell out LDA and GGA at least once in the introduction already.}. 
A natural choice of $H^0$ is thus the KS-Hamiltonian $H^\text{KS}(\bfk)$ within LDA or GGA, that contains 
the kinetic-energy operator, the external potential ($v_\text{ext}$), the Hartree potential ($v_\text{H}$), and the exchange-correlation (XC) 
potential  ($v_\text{\rm XC}$)
      \begin{equation}
           H^\text{KS}(\bfk) = - \frac{1}{2} \nabla^2 + v_\text{ext}(\bfk) + v_\text{H}(\bfk)
                    + v_\text{\rm XC}(\bfk)\, .
      \end{equation}
Next, we need to define $\Sigma^\text{loc}(i\omega)$ in Eq.~(\ref{eq:on_site_gf}). 
If we start from $H^\text{KS}(\bfk)$,   the ``on-site" self-energy becomes the difference between the dynamic, complex many-body exchange-correlation 
self-energy $\Sigma_\text{\rm XC}(\bfk,i\omega)$ and the KS XC potential, i.e.,
     \begin{align}
          \Sigma^\text{loc}(i\omega) & =  \frac{1}{N_{\rm 1.BZ}}\sum \limits ^{\rm 1.BZ}_{\bf k}
            \left[\Sigma_\text{\rm XC}(\bfk, i\omega) - v^\text{KS}_\text{\rm XC}(\bfk) \right] \nonumber \\
              & = \Sigma_\text{\rm XC}^\text{loc}(i\omega) - v_\text{\rm XC}^\text{loc} \, .
          \label{Eq:sigma_loc}
     \end{align}
Using Eqs.~(\ref{eq:on_site_gf}) and (\ref{Eq:sigma_loc}), we finally obtain
	\begin{align}
           &	G^{\mathrm{on \mhyphen site}}_{ij}(i\omega) = \nonumber \\
           &   \frac{1}{N_{\rm 1.BZ}}\sum \limits ^{\rm 1.BZ}_{\bf k} \left[
            (i\omega + \mu)S(\bfk) - H^\text{KS}(\bfk) - \Sigma^\text{loc}_\text{\rm XC}(i\omega) + v^\text{loc}_\text{\rm XC} \right]^{-1}\, .
% 	  \label{eq:on_site_gf}
	  \end{align}
Our scheme is thus free from any double-counting ambiguities, because the DFA XC-contribution that has to be subtracted is uniquely defined.

\subsubsection{Embedded Green's function}

In the DMFT formalism, a periodic system is viewed as a periodically repeated cluster 
    (here the unit or supercell) dynamically embedded into
     a self-consistently determined environment. The coupling between the embedded subsystem and its
     surrounding environment is described by a so-called bath Green's function ${\cal G}(i\omega)$,
     connecting the Green's function of the embedded cluster $G^\text{emb}(i\omega)$ and the
     local self-energy via
        \begin{equation}
             \left[{\cal G}(i\omega)\right]^{-1} = \left[G^\text{emb}(i\omega)\right]^{-1} + 
               \Sigma^\text{loc}(i\omega)    \, .
            \label{Eq:dmft_scf1}
        \end{equation}
     Here the local self-energy $\Sigma^\text{loc}(i\omega)$ is the same as introduced in Eq.~(\ref{Eq:sigma_loc}).
     The self-consistency condition of DMFT requires that the Green's function of the embedded cluster
     $G^\text{emb}(i\omega)$ equals the on-site Green's function as given in Eq.~(\ref{eq:on_site_gf}),
        \begin{equation}
           G^\text{emb}(i\omega)=G^\mathrm{on \mhyphen site}(i\omega)\, .
           \label{Eq:dmft_scf2}
        \end{equation}

    Alternatively, one can also use a so-called hybridization function $\Delta(i\omega)$ to describe the
    coupling between the 
    embedded cluster and its environment, which provides a more intuitive picture. $\Delta(i\omega)$ is closely 
    related to the bath Green's function
    ${\cal G}(i\omega)$, 
	\begin{eqnarray}
		\left[\mathcal{G}(\textit{i}\omega)\right]^{-1} &=& (\textit{i}\omega +\mu)S- H^{\rm cluster}_{\rm 0} -\Delta(\textit{i}\omega)\, .
		\label{Eq:hybrid_func}
	\end{eqnarray}
    In Eq.~(\ref{Eq:hybrid_func}) $H^{\rm cluster}_{\rm 0}$ is the Hamiltonian of the bare cluster describing the non-interacting unit
    cell i.e., without the $v^\text{KS}_\text{\rm XC}$ contribution and without the presence of the other atoms from neighboring unit cells (see Fig.~\ref{fig:dmft_concept}). This corresponds to the ``on-site" term of the Hamiltonian of the 
    periodic system, and in practice can be conveniently obtained from the $\bfk$-dependent Hamiltonian,
	\begin{eqnarray}
		H^{\rm cluster}_{\rm 0} &=& \frac{1}{N_{\rm 1.BZ}}\sum \limits ^{\rm 1.BZ}_{\textbf{k}}\left[ H^{\rm KS}(\textbf{k})-v^\text{KS}_\text{\rm XC}(\bfk)\right]\, .
	\end{eqnarray}

  Using eqs.~(\ref{Eq:dmft_scf1})-(\ref{Eq:hybrid_func}), we obtain the following expression for 
the Green's function of the embedded cluster
	\begin{align}
		&\biggl[G^{\rm emb}(i\omega)\biggr]^{-1}_{ij} = \nonumber \\
		&\biggl[(i\omega +\mu)S - H^{\rm cluster}_{\rm 0} -  \Sigma^{\rm loc}[G^{\rm emb}](i\omega) -\Delta(i\omega)\biggr]_{ij}\, .
		\label{Eq:embed_gf}
	\end{align}
Here we have explicitly indicated that the local self-energy is a functional of the embedded Green's function.
Thus eq.~(\ref{Eq:embed_gf}) has to be solved self-consistently, which corresponds to the inner
loop of Fig.~\ref{fig:embedding_loop}. The functional dependence of $\Sigma^{\rm loc}(i\omega)$ on $G^{\rm emb}(i\omega)$ is given by the actual approximation for the localized region, which will be the topic of next section. However, already here we see that our RDMFE approach lends itself to those advanced electronic-structure methods that can be expressed by (self-consistent) Green's functions.

Another point we would like to emphasize is the choice of the cluster overlap matrix $S$ in eq.~(\ref{Eq:hybrid_func}). 
% This affects only the hybridizaton function $\Delta(i\omega)$, but not the entire bath Green's function ${\cal G}(i\omega)$, which 
% finally determines the coupling effect between the embedded Green's function and the bath. In this sense, the different 
% choice of $S$ will not affect the physical results here.
%     The proper definition of the overlap matrix more delicate.
We found  when updating the chemical potential of the cluster in the inner loop that we needed to define the cluster overlap matrix as  $ S= \biggl[\frac{1}{N_{\rm 1.BZ}}\sum \limits ^{\rm 1.BZ}_{\bf \textbf{k}} S^{-1}(\textbf{k}) \biggr]^{-1}$ (and not simply $ S= \frac{1}{N_{\rm 1.BZ}}\sum \limits^{\rm 1.BZ}_{\bf k} S(\textbf{k})$) as done by Kotliar \textit{et al.} \cite{Kotliar/etal:2006} to enforce 
the correct asymptotic behavior of $\Delta(i\omega)$, i.e. $\lim\limits_{\omega\longrightarrow\infty}\Delta(\omega)\longrightarrow0$.

\subsection{The local self-energy}
     So far we had not specified the approximation for the local self-energy in eq.~(\ref{Eq:embed_gf}). In our scheme the choice for 
     $\Sigma^{\rm loc}[G^{\rm emb}]$ can be quite flexible. In other words, we could use any approximation that goes beyond LDA and GGAs. However, our framework lends itself to Green's-function-based approaches. This includes density-matrix and density-based approaches, because both quantities can easily be extracted from the Green's function.      Below we report on two different examples, namely hybrid density functionals that mix a fraction of exact-exchange with GGA semi-local exchange \cite{Becke:1993,Perdew/Burke/Ernzerhof:1996,Heyd/Scuseria/Ernzerhof:2003} and the $GW$ approximation. \cite{Hedin:1965} 
     In practice, we could also go beyond $GW$, e.g., by including the screened second-order exchange (SOSEX) self-energy that was developed recently. \cite{Ren/etal:2015}  

We here use the PBE hybrid functional family (PBEh) \cite{Ernzerhof/etal:1999}, whose most prominent functional is PBE0. \cite{Adamo/etal:1999} We will also use the short-ranged range-separated hybrid functional family by Heyd, Scuseria and Ernzerhof (HSE). \cite{Heyd/Scuseria/Ernzerhof:2003} In PBEh the local self-energy in 
eq.~(\ref{Eq:sigma_loc}) is given by
	\begin{align}
		 \Sigma^{ \rm loc}_{\rm PBEh} (\alpha) &  =  
		\left[\alpha\Sigma^{\rm loc}_{\rm X}+ (1-\alpha) v^{\textrm{loc}}_{\rm X}+ 
              v^{\textrm{loc}}_{\rm C}\right] - v^{\textrm{loc}}_\textrm{\rm XC}  \nonumber \\
              & = \alpha \left(\Sigma^{\rm loc}_{\rm X} - v^{\textrm{loc}}_{\rm X} \right) .
          \label{Eq:PBE0}
	\end{align}
In Eq.~(\ref{Eq:PBE0}), $v^{\textrm{loc}}_{\rm X}$ is the ``on-site" part of the GGA exchange,  and
     $\Sigma^{\rm loc}_{\rm X}$ is the exact-exchange matrix given by
	\begin{eqnarray}
		\left[ \Sigma_{\rm X}^{\rm loc}\right]_{ ij} =  \sum \limits_{ k, l} \langle ik 
		|  lj \rangle  n^{\textrm{emb}}_{kl}, 
		\label{loc_self_enrg_exx}
	\end{eqnarray}
% \PR{please be consistent in your subscripts. Sometimes you denote exchange by lower case \emph{x} and sometimes by upper case \emph{\rm X}. Pick one. You also sometimes set x or xc in roman font and other times in italics. Again, pick one. I suggest to pick roman font.}
where $\langle ik|lj\rangle$ are two-electron four-orbital integrals, and $ n^{\textrm{emb}}_{kl}$ is
the density matrix of the embedded cluster which can be obtained from the embedded Green's function
	\begin{eqnarray}
		 n^{\rm emb}_{ij} = - \frac{i}{2\pi}\int  \! d\omega  \, G^{\rm emb}(i\omega)_{ ij}e^{i \omega \tau^{+}}.
         \label{Eq:denmat}
	\end{eqnarray}
The two-electron Coulomb repulsion integrals are evaluated using the resolution of identity (RI) technique in FHI-aims as  documented in Ref.~\onlinecite{Ren/etal:2012}. 
The PBE0 functional is obtained for $\alpha$=0.25. \cite{Adamo/etal:1999}

The extension to an HSE type self-energy is straightforward. In HSE, a range-separation parameter is introduced that cuts off the exact-exchange contribution at long distances. 
The range is controlled via the screening parameter $\gamma$ 
% \PR{can we give the screening parameter a different symbol so that it doesn't get confused with the frequency argument of the Green's function? What about $\gamma$?}, 
so that the local exchange self-energy becomes
	\begin{align}
	  \Sigma_{\rm X}^{\rm loc}(\gamma) = \Sigma_{\rm X}^{\rm loc,SR}(\gamma) +\Sigma_{\rm X}^{\rm loc,LR}(\gamma),
	\end{align}
with SR and LR denoting the short and long-range part, respectively. 
If we now replace $\Sigma_{\rm X}^{\rm loc,LR}(\gamma)$ by $v_{\rm X}^{\rm loc,LR}$ and introduce the $\alpha$ parameter again,  the local HSE self-energy assumes the following form
	\begin{align}
	  \Sigma_{\rm HSE}^{\rm loc}(\alpha,\gamma) = \alpha \left(\Sigma^{\rm SR,loc}_{\rm X}(\gamma) - v^{\textrm{SR,loc}}_{\rm X}(\gamma) \right)
	\end{align}
% \PR{This expression cannot be correct for HSE. 
% It should be $\Sigma_{\rm x}^{\rm HSE}(\omega)=v_{\rm x}^{\rm PBE}-\alpha v_{\rm X}^{\rm PBE,SR}(\omega)+\alpha \Sigma_{\rm x}^{\rm HF,SR}(\omega)$. 
% If we now subtract $v_x$ to obtain $\Sigma^{\rm loc}$ we obtain $\Sigma_{\rm HSE}^{\rm loc}(\alpha,\omega) = \alpha \left(\Sigma^{\rm SR,loc}_{\rm X}(\omega) - v^{\textrm{SR,loc}}_{\rm X} \right)$. In other words, the PBE exchange potential you subtract is short-ranged. 
% I hope this is implemented!}

Furthermore, we employ the $GW$ approximation for the local self-energy. Here, the computation of
the $GW$ self-energy for a given input embedded Green's function follows  the self-consistent $GW$ implementation
for finite systems in FHI-aims. \cite{Ren/etal:2012,Caruso/etal:2013} On the imaginary time axis, the $GW$ self-energy for the embedded cluster is obtained as
	\begin{equation}
		[\Sigma^{\rm loc}_{\rm XC}( i \tau)]_{ij}=
		\frac{i}{2\pi} \sum\limits_{ lk  \mu\nu} M^{\mu}_{ik} M^{\nu}_{lj} G^{\rm emb}_{kl}(i\tau)[W^{\rm loc}( i \tau)]_{ \mu \nu}\, .
		\label{Eq:loc_self_enrg_corr}
	\end{equation}  
% \PR{The way eq.~\ref{Eq:loc_self_enrg_corr} is written, i.e. G*W, it still contains the exchange contribution. It thus gives $\Sigma_{xc}$ and not $\Sigma_c$. Otherwise it would have to be G * (W-v). Furthermore, the sum contains an index $m$, which does not show up in the argument of the sum! Please make sure that all formulas are correct!}
Here $\mu,\nu$ indices refer to the auxiliary basis set used to expand the screened Coulomb interaction
$W^{\rm{loc}}$ in the RI approach \cite{Ren/etal:2012,Caruso/etal:2013}. Furthermore $M^{\mu}_{ik}$ are the 3-index coefficients obtained as, 
  \begin{equation}
       M^{\mu}_{ik} = \sum_v (ik|\mu) V_{\mu\nu}^{-1/2}\, ,
  \end{equation}
where
  \begin{equation}
     (ik|\mu)=\int d\bfr d\bfrp \frac{\phi_i(\bfr)\phi_k(\bfr)P_\mu(\bfrp)}{|\bfr-\bfrp|}\, ,
  \end{equation}
and 
  \begin{equation}
     V_{\mu\nu} = \int d\bfr d\bfrp \frac{P_\mu(\bfr)P_\nu(\bfrp)}{|\bfr-\bfrp|}
  \end{equation}
with $\{P_\mu(\bfr)\}$ being the auxiliary basis functions. 
For $W$ we thus obtain
	\begin{eqnarray}
   W^{\rm loc}_{ \mu \nu}( i \omega)= \sum_{\alpha} V_{\mu\alpha} [1-\Pi^{\rm loc}( i\omega)]^{-1}_{\alpha\nu} \, 
	\end{eqnarray}
where $\Pi^{\rm loc}( i\omega)$ the irreducible polarisability,
whose Fourier transform in the time domain is directly
determined by the embedded Green's function
    	\begin{eqnarray}
		\Pi^{\rm loc}_{ \mu \nu}( i \tau)= -i\sum\limits_{ ij lm} M^{\mu}_{il} M^{\nu}_{jm} G^{\rm emb}_{ij}(i\tau) G^{\rm emb}_{lm}(-i\tau).
	\end{eqnarray} 
% \PR{Shouldn't there be a $-i$ in front of the sum? P=-iG*G!}
% \PR{Also, do we need to address how the overlap matrix is handled? I seem to recall that Wael had it in the wrong place, before we double checked with Fabio again. To prevent the reader from making the same mistake, we could clarify this point here.}
% \Wael{If we would like to address how the overlap matrix is handled, we should do it where we define the lattice GF. I am not sure if I still have to do it...} \XR{Anyhow, I now explained what $M^{\mu}_{il}$ is in the text ...}

      \subsection{The self-consistency loops}
\begin{figure}[htp]
	    \centering
 \includegraphics[scale=0.25]{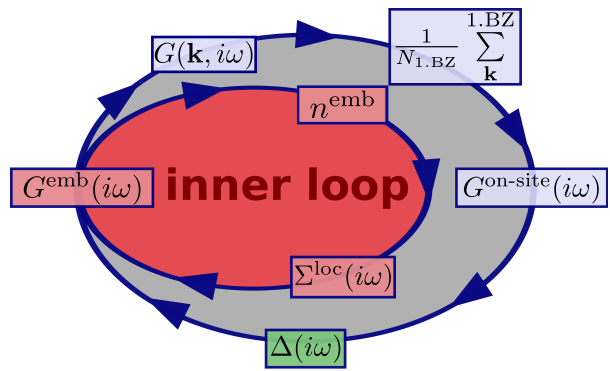}
 \caption{The embedded (Eq. (\ref{Eq:embed_gf})) and the on-site (Eq. (\ref{eq:on_site_gf})) Green's functions define two Dyson equations that form two nested loops. 
 The two loops are iterated until self-consistency is reached.}\label{fig:embedding_loop}
\end{figure}

    In our formalism eqs. (\ref{Eq:embed_gf}) and (\ref{loc_self_enrg_exx}) or (\ref{Eq:loc_self_enrg_corr}) define an additional inner self-consistency loop for the local self-energy  as depicted in Fig.~\ref{fig:embedding_loop}. 
    Good convergence is achieved by a linear mixing
	\begin{eqnarray}
		\Sigma_{n +1}^{\rm loc}= \lambda\Sigma_{ n}^{\rm loc} + (1-\lambda)\Sigma_{n-1}^{\rm loc},
	\end{eqnarray}
    with a mixing parameter $\lambda=0.5$. More advanced mixing schemes could be implemented as well, but we found that linear mixing works well for the  examples presented in this work. 
    When the inner-loop reaches convergence we feed the resulting $\Sigma^{\rm loc}$ back into the on-site GF and iterate the main-loop further using the same mixing as for the inner-loop. 
%     \PR{Similar? Or identical? Do you have one or two mixing parameters in your DMFT scheme in practice? This would be important to know for a user.}.

\begin{figure}[htp]
   \begin{center}
   \includegraphics[scale=0.25]{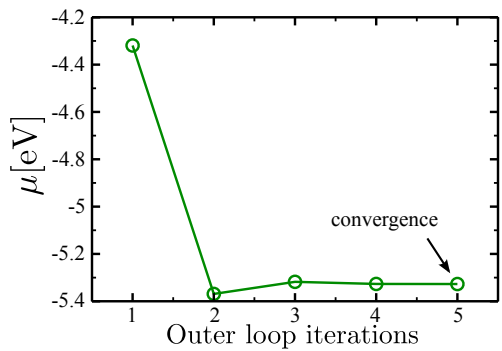}
   \caption{Typical change in the chemical potential during the self-consistency cycle for a bulk Si calculation.
   Convergence is reached after 5 iterations of the outer loop. Already at the second iteration the chemical potential is close to its converged value.}
   \label{fig:chem_pot}
   \end{center}       
   \end{figure}

    Finally it is worth mentioning, that the on-site Green's function as defined in eq.~(\ref{eq:on_site_gf}) requires that our $\Sigma^{\rm loc}$ in the on-site Green's function in the $0$-th iteration should be
    $	\Sigma^{\rm loc}_{\rm 0}=V^{\rm loc}_{\rm XC}$.
Figure~\ref{fig:embedding_loop} shows a sketch of the embedding scheme as described above. 
During the self-consistency cycle we compute the particle number $N_{\mu}$, a quantity that is obtained from the
embedded Green's function via 
	\begin{align}
	  N_{\mu} = -\frac{i}{2\pi}\text{Tr} \int d\omega G^{\rm emb}_{ij}(\omega,\mu)e^{-i\omega 0^{+}}.
	\end{align}
To ensure particle number conservation, we need to update the electron chemical potential every time we receive a converged self-energy from the inner-loop. 
For the present test cases, we found that the change in the chemical potential is relatively small, as demonstrated in Fig.~\ref{fig:chem_pot} for bulk silicon (Si). However, we expect it to be more important for 
metallic systems. 

% \PR{Ok, this paragraph is a bit imprecise! First, how is the particle number $N$ obtained? Please give an equation. Second, you say \emph{we should} update the chemical potential in a third loop, but the presence of the word \emph{should} implies that you \emph{don't}. Then the question is, why not? I think the chemical potential is important enough to warrant further attention! Especially since we battled with it quite a bit during our development. Also, when it comes to metals, the chemical potential is essential and if you can get your RDMFE converged without the 3rd chemical potential loop that's a major achievement. Furthermore, the particle number $N$ is an important quantity that you monitor to guarantee that your scheme is behaving properly. This needs to be mentioned.}

\subsection{\label{sec:total_energy}Total energy calculation}
   Once self-consistency in the embedding scheme is reached, we can compute the total energy of the entire systems (the embedded cluster plus the environment) using the converged lattice Green function $G_{ij}(\textbf{k},\omega)$. The actual total-energy expression depends on the chosen methodology used in the embedded region.
For hybrid density functionals we have
     \begin{align}
          E^{\rm hyb}_{\rm tot} = & \frac{1}{N_{\rm 1.BZ}}\left[ \sum \limits_{\textbf{k}} \sum \limits_{i,j} 
                   t_{ji}(\textbf{k})  n_{ij}(\textbf{k})\right] \nonumber \\
          &  +  E_{\rm es}[n] + E^{\rm hyb}_{\rm xc}[n_{ij}]  \, ,
       \label{Eq:hybrid_Etot}
     \end{align}
where 
    \begin{equation}
      t_{ji}(\textbf{k})=\sum_{\bfR} 
        \langle \phi_j(\bfr)|-\frac{\nabla^2}{2}|\phi_i(\bfr-\bfR)\rangle e^{i\bfk \cdot \bfR}
    \end{equation}
is the matrix form of the kinetic energy operator,
 $E_{\rm es}$  the electrostatic (Hartree plus external) energy,
and $E^{\rm hyb}_{\rm xc}$ the XC energy. In Eq.~(\ref{Eq:hybrid_Etot}) $n_{ij}$ is the $\bfk$-dependent global density matrix,
    \begin{equation}
        n_{ij}(\textbf{k})= \int \frac{d\omega}{2\pi } G_{ij}^{\rm lat}(\textbf{k},i\omega)e^{-i\omega 0^{+}}\, ,
    \end{equation}
and $n$ is the electron density obtained from $n_{ij}(\textbf{k})$. 
We note that Eq.~(\ref{Eq:hybrid_Etot})
is the exact total-energy expression for the hybrid density functional, and the only approximation is that
the density matrix $n_{ij}(\textbf{k})$ (and hence electron density $n$) is obtained from the RDMFE scheme 
and not from a periodic hybrid functional calculation. 

However, Eq.~(\ref{Eq:hybrid_Etot}) cannot be directly applied 
since evaluating $E^{\rm hyb}_{\rm xc}$ as a functional of the $\bfk$-dependent density matrix $n_{ij}(\textbf{k})$ 
requires the computation of the exact-exchange energy for the entire periodic system, which is exactly what 
we are trying to avoid here. Instead of evaluating $E^{\rm hyb}_{\rm xc}[n_{ij}(\bfk)]$ in full, we thus only compute the change of $E^{\rm hyb}_{\rm xc}$ with respect to the local or semi-local (LDA or GGA) energy in the embedded region.  
This is the main approximation of our approach, which is consistent with the spirit of the local self-energy correction  in the RDMFE scheme, 
and is suggested by the near-sightedness of the XC energy of a bulk system (although the exact-exchange energy is probably not the most near-sighted self-energy we could have chosen). \cite{Kohn:1995}

The Hartree energy depends on the electron density in a highly non-local way and it is questionable if a local treatment can be applied to the Hartree energy at all. Therefore, for simplicity, we omit possible changes in the Hartree and the external energy for now, assuming that the electron density given by the local or semi-local approximation is already sufficient. 

Finally, we are left with the kinetic energy term which also changes when moving from local or semi-local to hybrid functionals. For consistency, kinetic and XC energy should be taken together. In our scheme, we evaluate the changes of the kinetic and XC energy caused by the local self-energy correction within the embedded region.

Based on the above considerations, we propose the following approximate total-energy expression for embedded hybrid functional calculations
    \begin{align}
      E_\text{tot}^\text{hyb} \approx & E_\text{tot}^\text{KS} + \left[\sum_{ij}t_{ji}(n_{ij}^{\rm emb} - n_{ij}^\text{KS})\right] \nonumber  \\
        & +  E_\text{XC}^{\rm hyb}[n^\text{emb}_{ij}] -E_{\rm XC}^\text{KS}[n^\text{KS}] \, ,
      \label{Eq:Etot_hyb_approx}
    \end{align}
where $n^\text{emb}_{ij}$ is the embedded density matrix as defined in Eq.~(\ref{Eq:denmat}), and
$n_{ij}^\text{KS}$ is the ``on-site" density matrix of KS-LDA/GGA calculations. $n^\text{KS}(\bfr)$ is obtained from the on-site KS density matrix
    \begin{equation}
        n^\text{KS} (\bfr)=  \sum_{ij} \phi_i(\bfr) n_{ij}^\text{KS} \phi_j(\bfr)\, ,
    \end{equation}
and $E_{\rm XC}^\text{KS}[n^\text{KS}]$ is thus restricted to the embedded region.

For $GW$ we can proceed in an analogous fashion
   \begin{align}
        E^{GW}_{\rm tot} = & \frac{1}{N_{\rm 1.BZ}}\left[\sum \limits_{\textbf{k}} \sum \limits_{i,j} 
                 \left[t_{ji}(\textbf{k})  n_{ij}(\textbf{k})\right]\right] \nonumber \\
              & + E_{\rm es}[n] + E^{GW}_{\rm XC}[G^{\rm lat}_{ij}]  \, ,
   \end{align}
where
  \begin{equation}
    E^{GW}_{\rm XC}[G^{\rm lat}_{ij}] = \frac{1}{2}\sum_{\bfk} \frac{1}{N_{\rm 1.BZ}}\int \frac{d\omega}{2\pi }\Sigma_{ji}(\textbf{k},\omega) G_{ij}(\textbf{k},\omega)e^{-i\omega 0^{+}}
  \end{equation}
following directly from the Galitskii-Migdal (GM) formula. \cite{Galitskii/Migdal:1958}
% \XR{Here we assume that the spin summation is already included in $G_{ij}$ and $n_{ij}$}.
Similar to the hybrid functional case, we will not take the full $\bfk$ dependence in $E^{GW}_{\rm XC}[G^{\rm lat}_{ij}(\bfk,i\omega)]$ into account. Instead we adopt the same philosophy as before and make a local approximation
    \begin{align}
      E_\text{tot}^{GW} \approx & E_\text{tot}^\text{KS} + \sum_{ij}t_{ji}(n_{ij}^{\rm emb} - n_{ij}^\text{KS})
          \nonumber \\
        &  +  E_\text{XC}^{GW}[G^\text{emb}_{ij}(i\omega)] -E_{\rm XC}^\text{KS}[n^\text{KS}] \, ,
      \label{Eq:Etot_hyb_approx}
    \end{align}
where
   \begin{equation}
         E^{GW}_{\rm XC}[G^{\rm emb}_{ij}] = \frac{1}{2}\int \frac{d\omega}{2\pi i}\Sigma_{ji}^{\rm loc}(i\omega) 
         G_{ij}^{\rm emb}(i\omega)e^{-i\omega 0^{+}} \, .
         \label{gw_xc_contribution_GM}
   \end{equation}

To summarise this part, in RDMFE the total energy of the entire system can in principle be obtained from the lattice Green's function. However, in practice, approximations are needed to make the problem tractable.  The expressions for hybrid functional and $GW$ calculations proposed above are consistent with the local nature of the self-energy approximation in RDMFE, but their performance needs to be checked in practical calculations. Future work needs to revisit total energy calculations in RDMFE.

% \XR{One more rigorous approach is to use the density matrix $n(\bfr, \bfrp)$ in real space to compute the 
% exact-exchange energy for hybrid functionals, but the integration over $\bfr$ and $\bfrp$ is restricted within 
% the embedded region (otherwise it goes back to periodic hybrid functional calculations).  This treatment will be 
% different from what is proposed above which only uses the on-site $n_{ij}$, since $n(\bfr,\bfrp)$ accounts for 
% contributions from basis functions from neighboring unit cells, and the density obtained from it $n(r)=n(r,r)$ 
% is the correct density of the periodic system. The same thing can be done for $GW$. I think this approach is 
% conceptually more correct, but the numerial implementation is much more involved. I am afraid that Wael won't 
% have time to do it for his PhD.  I hope that one day we can come back to do things in this more correct way. }

\section{\label{sec:comp} Computational details}
\begin{figure}[htp]
    \begin{center}
    \begin{tabular}{cc}
     {\includegraphics[scale=0.25]{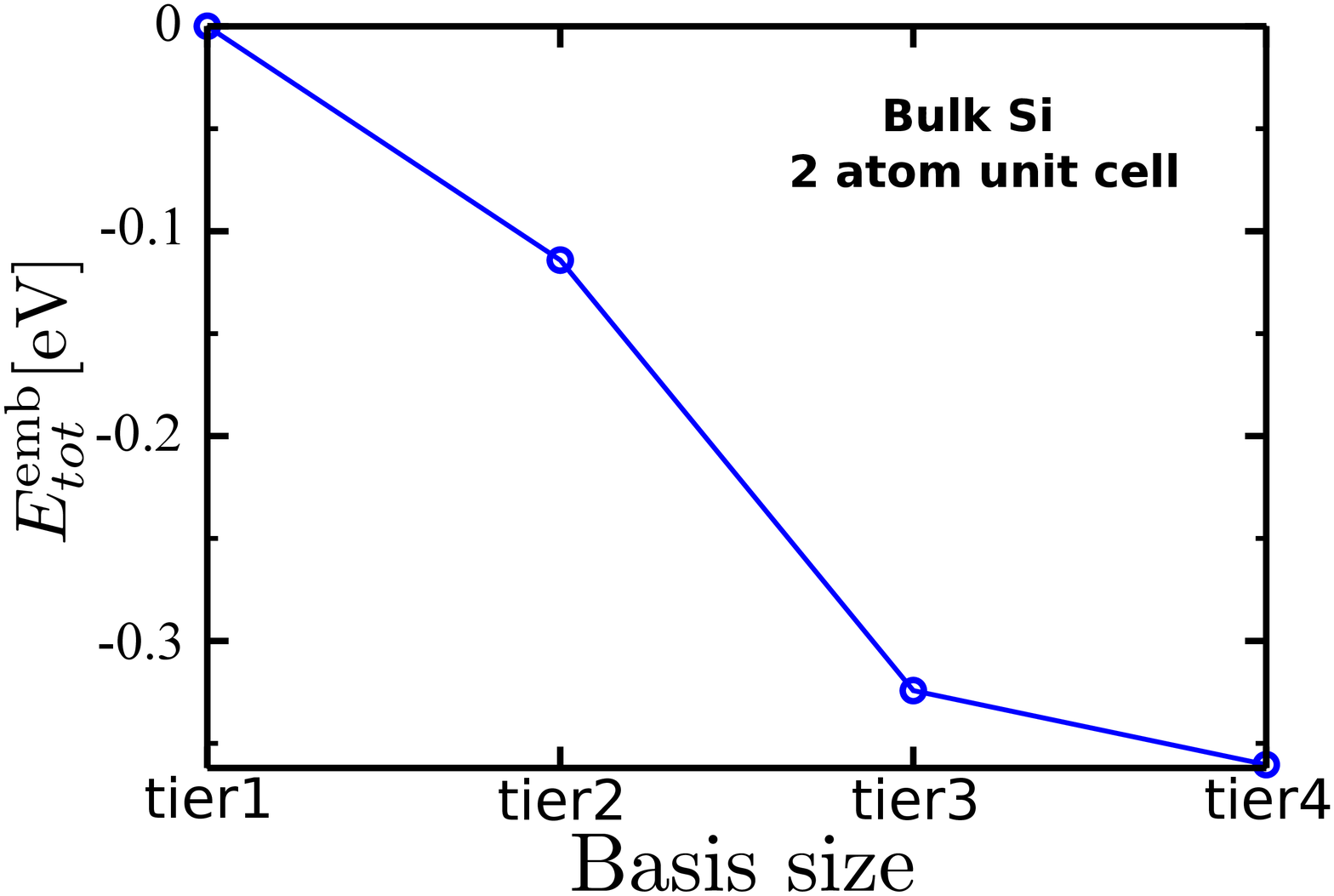}}\\
      {\includegraphics[scale=0.25]{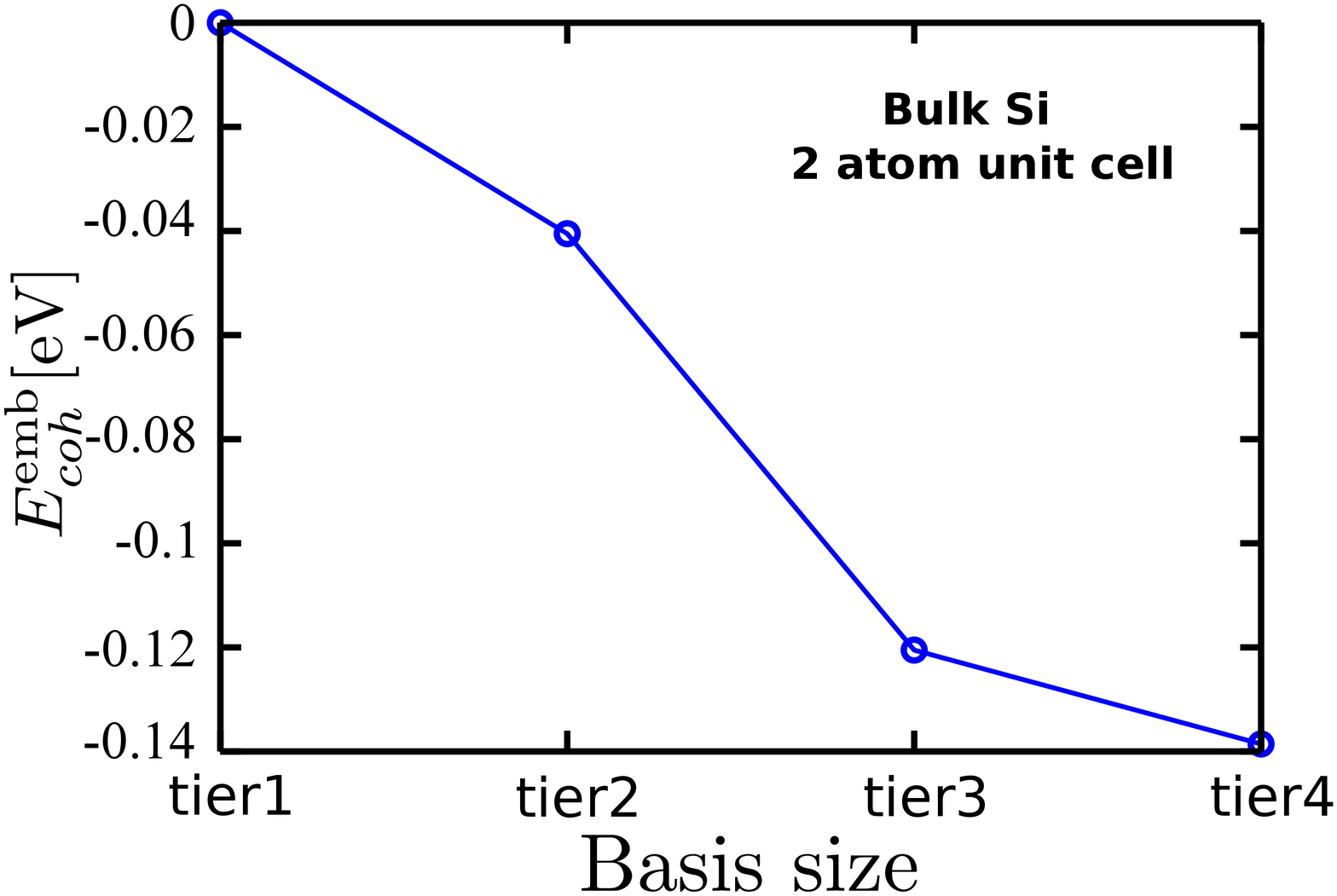}}
     \end{tabular}
     \caption{Embedded PBEh total energy  (upper panel) and cohesive energy (lower panel) for bulk Si with a 2 atom unit cell as function of the basis     size. The energy zero is set at the value of the \textit{tier} 1 basis.
     }\label{fig:Etot_vs_basis}
    \end{center}           
   \end{figure}

We used tight FHI-aims integration grids for all our RDMFE calculations. For the embedded PBEh and HSE self-energy we used the \textit{tier} 1 basis set.
Figure \ref{fig:Etot_vs_basis} shows the embedded PBEh total and cohesive energies with increasing basis set size.
The lattice Green's functions were represented on a logarithmic frequency grid with 40 points. The total energy calculations for 2 and 8 atom silicon unit cells 
and the density of states (DOS) calculations for 8 and 16 atom unit cells were performed on a $4\times 4\times 4$ \textbf{k}-point grid, which we increased 
to  $10\times 10\times 10$ for DOS calculations in the 2 atom unit cell. 
\textit{GW} calculations were performed in a \textit{tier} 3 basis set with 40 frequency/time points in the inner loop and the same number of \textbf{k}-points as for PBEh. 
The linear mixing parameter $\lambda$ was fixed to $0.5$, which gave reasonably fast convergence. 
The periodic PBE and PBE0 reference calculations were performed using the \textit{tier} 1 basis set and a  $12\times 12\times 12$ \textbf{k}-mesh.

Densities of states are obtained in two different ways. In PBEh and HSE we obtain a self-energy that defines a converged \textbf{k}-dependent embedded Hamiltonian via the 
converged lattice Green's function  once the self-consistency cycle is converged. For the PBEh self-energy we can directly diagonalize the embedded Hamiltonian 
$H^{\rm embed}(\textbf{k}) = H^{0}(\textbf{k}) + \Sigma^{\rm loc}_{\rm PBEh}$ 
at each \textbf{k}-point, which yields \textbf{k}-dependent eigenvalues and eigenstates. The resulting density of states (DOS) is
$n(\epsilon(\bfk))=\sum\limits_{\nu}\delta (\epsilon(\bfk)-\epsilon_{\nu}(\bfk))$, where $\nu$ labels the eigenstates of $H^{\rm embed}(\textbf{k})$.
To make the comparison to  experiment easier, we introduce a Gaussian broadening
   		      \begin{align}          
   		        g_{\nu} (\tilde{\epsilon}) = \int d\textbf{k}\text{ }exp\left(-\frac{\tilde{\epsilon}-\epsilon_{\nu}(\bfk)}{\sqrt{2}\sigma}\right)^{2}
                        \label{Eq:dos_gaussian}
   		      \end{align}
     to obtain the DOS 
    $N(\tilde{\epsilon}) = \sum\limits_{\nu} g_{\nu} (\tilde{\epsilon})$. 
%     \PR{Does the error function give us Gaussians? I thought a Gaussian looked like $e^{-\frac{(\tilde{\epsilon}-\epsilon_{\nu}(\bfk))^2}{2\sigma^{2}}}$}
In this work we use a Gaussian broadening of $\sigma=0.2$ eV.
%  \XR{ I don't understand why we use the error function instead of gaussian itself in
% the DOS calculation.} \PR{I see, Xinguo had the same question as I ;-)} 
% \Wael{Yes, but this is how it is implemented in FHI-aims, and since we want to compare with its DOSs I just took the exact same formula.}
% \XR{I think there is a misunderstanding of the source code. It uses the difference of the error function at two neighboring energy points to represent Gaussian. No sure why it does so, but it is not the error function itself.}

For the \textit{GW} self-energy, the spectrum at each \textbf{k}-point is directly given by the Green's function as
   		      \begin{align}           
   		        A(\textbf{k},\omega) = -\frac{1}{\pi} \text{Tr} \{\text{Im}G^{\rm lat}(\textbf{k},\omega)\}.
   		        \label{eq:k_spectral_func}
   		      \end{align}
To determine $G^{\rm lat}(\textbf{k},\omega)$ on the real-frequency axis, we analytically continue the self-energy from the imaginary to the real axis. 
In practice, we fit a two-pole model, that has proven to work very well for the systems we tested, to each matrix element of the self-energy\cite{Rojas/Godby/Needs:1995,Ren/etal:2012}
   		      \begin{align}           
   		        \Sigma(i\omega) \approx \sum\limits^{2}_{n}\frac{\alpha_{n}}{i\omega-\beta_{n}},
		        \label{eq:twopolemodel}
   		      \end{align}
where $\alpha_{n}$ and $\beta_{n}$ are complex fitting parameters. 
We then evaluate eq.~(\ref{eq:twopolemodel}) for real frequencies and solve Dyson's equation for $G^{\rm lat}(\textbf{k},\omega)$. 
The spectral function subsequently follows from a \textbf{k}-summation $A(\omega)=\sum\limits_{\textbf{k}}A(\textbf{k},\omega)$, 
which we convolute with Gaussians as
   		      \begin{align}           
   		         \tilde{A}(\tilde{\omega})= \int d\omega e^{-\left(\frac{\tilde{\omega}-\omega}{\sqrt{2}\sigma}\right)^{2}}A(\omega),
   		      \end{align}
with a broadening that we choose to be $\sigma=0.01$ eV to obtain a DOS $\tilde{A}(\tilde{\omega})$ that we can compare with experiment.

% \XR{I don't quite understand why one needs two different ways to compute the DOS, and why the broadening parameters
% are so different (0.2 eV vs 0.01 eV)}
% \XR{I am wondering how the gaussian convolution is done here. 
% I suppose it is different from Eq.~\ref{Eq:dos_gaussian}?} \PR{Can you give the final formula? How do you insert the Gaussians in $A(\omega)=\sum\limits_{\textbf{k}}A(\textbf{k},\omega)$?}

\section{\label{sec:results} Results}
Having introduced the concept of RDMFE and our implementation in the previous sections, we now turn to benchmark calculations for hybrid functionals, for which we have an independent, periodic reference in FHI-aims \cite{Levchenko/etal:2015}. Then we present self-consistent $GW$ calculations for which such a periodic reference does not yet exist in FHI-aims. We choose bulk Si as test system since it is a reliable and well studied reference case.
% \PR{Let's focus on Si for now, but it would be very nice, if we could include at least one metal in this paper!}

 \subsection{Density of states and band structures}

\begin{figure}[htp]
   \begin{center}
   \includegraphics[scale=0.3]{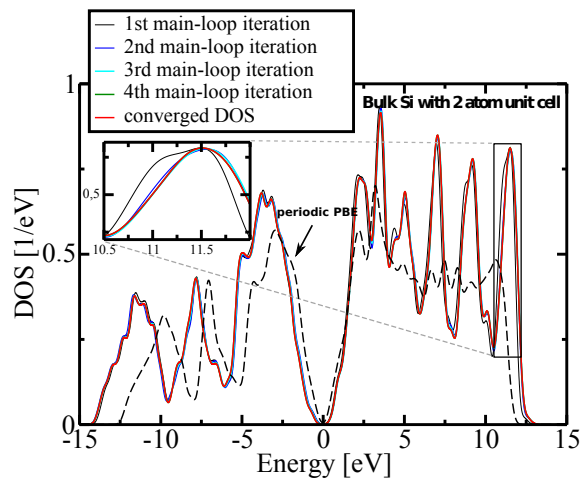}
   \caption{DOS comparison at each iteration of the main-loop. Convergence was achieved after 5 main-loop iterations.}\label{fig:dos_comp_iter}
   \end{center}
    \end{figure}

We begin our benchmark tests by calculating the DOS at each iteration to investigate its evolution with each embedding cycle. Figure~\ref{fig:dos_comp_iter} shows the DOS at different iterations of the outer loop for a 2 atom unit cell of silicon. 
 We observe that the largest change occurs at the first iteration when moving from PBE to our embedded PBE0 DOS. For subsequent iterations the DOS changes are  much smaller.
%    \PR{If your argument is that PBE0 does not differ too much from PBE and therefore the DOS converges quickly with iteration, then you should definitely show the same plot for GW! I think the main argument here is that the biggest change occurs already at the 1st iteration (i.e. when going from PBE to PBE0$_{n=1}$). I think Xinguo has also pointed this out and asked you to include the zeroth iteration, i.e. PBE, for exactly that reason!}
   \begin{figure}[htp]
   \begin{center}
   \includegraphics[scale=0.3]{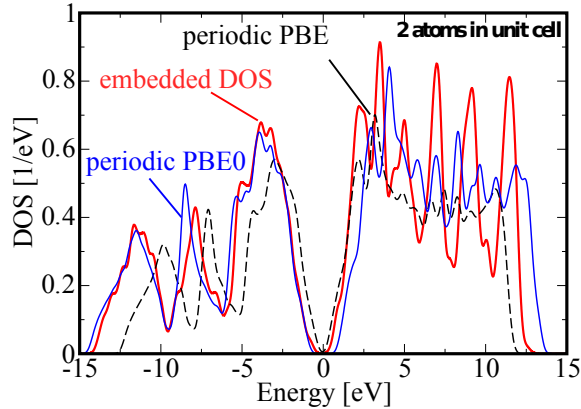}
   \caption{Comparison of the embedded DOS, the periodic PBE and the periodic PBE0 DOS for a 2 atom unit cell. 
%    \PR{The ``periodic PBE0'' reference curve is different in all panels. It looks different in Fig.~\ref{fig:dos_comp_2atom} and Fig.~\ref{fig:dos_comp_8atom_16atom} and even in \ref{fig:dos_comp_8atom_16atom} it doesn't agree in the two panels. Matthias also commented on that and the fact that there are still sub-peaks in the DOS that would disappear with a higher k-point sampling. Could you please increase the k-point sampling for the periodic PBE0 DOS calculation, so that we get a smooth PBE0 reference curve and then you use that as reference in all plots. Please do the same with the PBE curve. }
   }\label{fig:dos_comp_2atom}
   \end{center}
    \end{figure}
     \begin{figure}[htp]
   	    \centering
      \begin{tabular}{cc}
  {\includegraphics[scale=0.3]{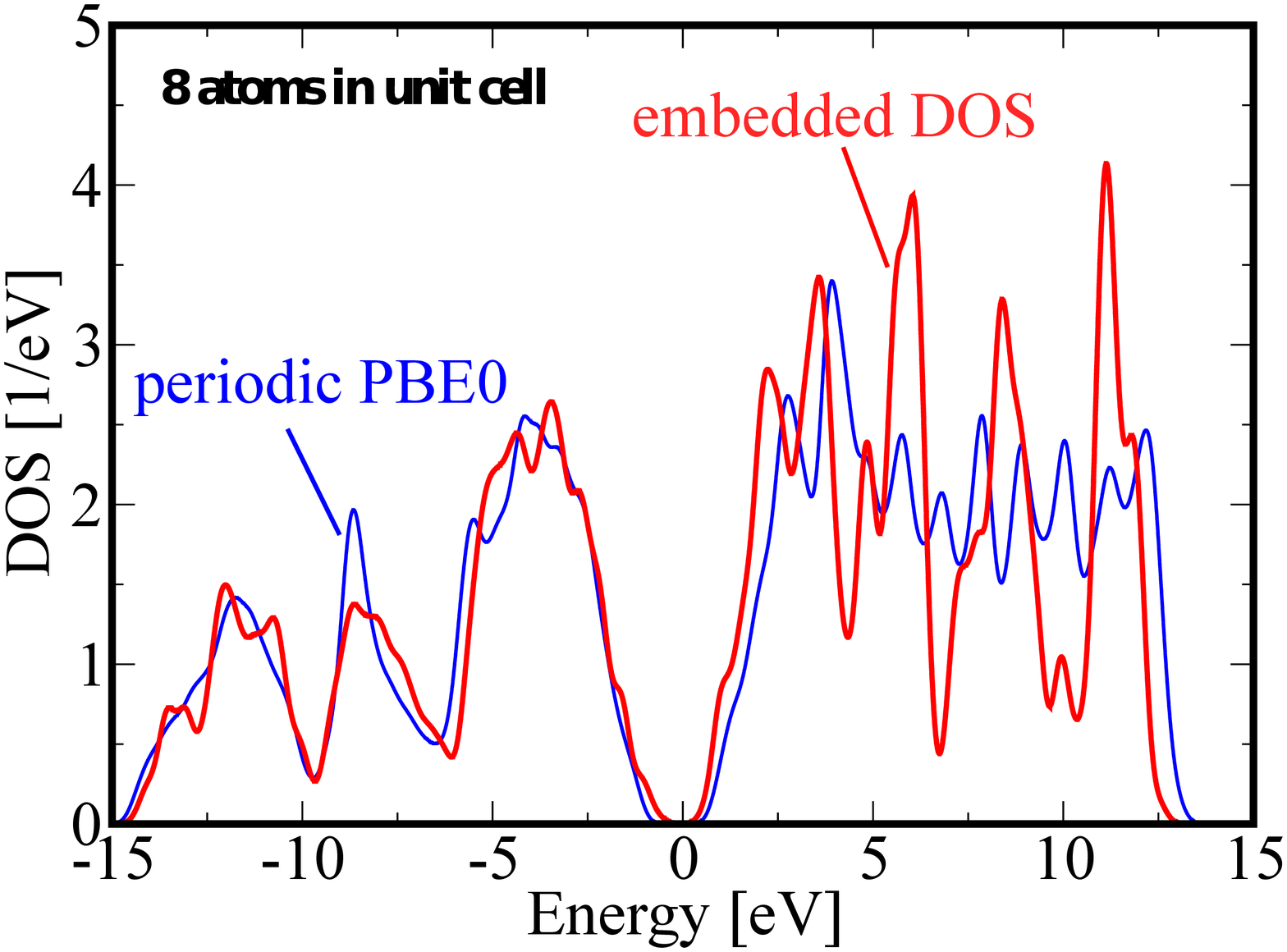}}\\
     {\includegraphics[scale=0.3]{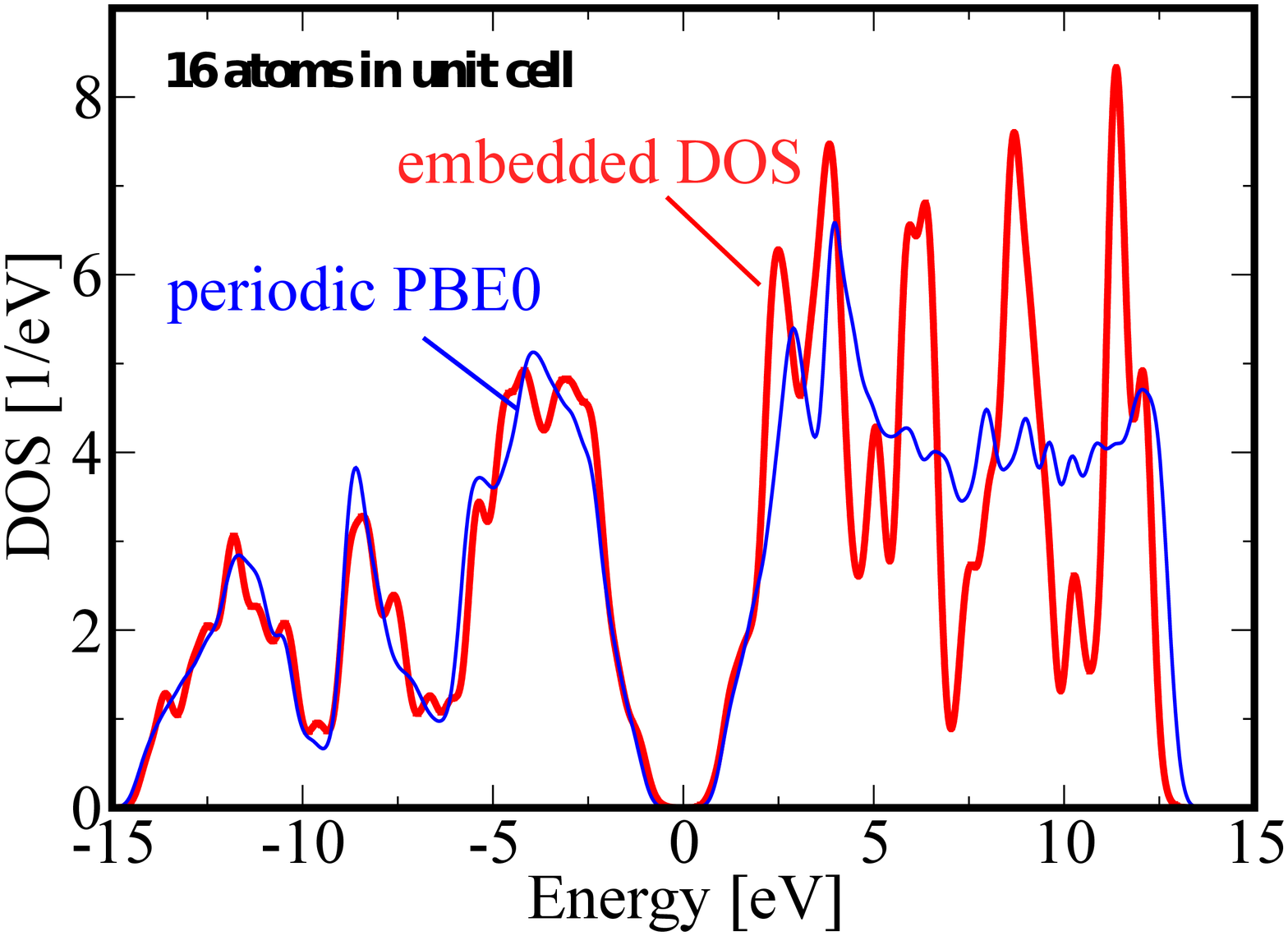}}
      \end{tabular}
   \caption{Comparison of the embedded DOS, with the periodic PBE0 DOS for an 8 atom unit cell 
   (upper panel) and a 16 atom unit cell (lower panel). 
%    \PR{Are you using the same broadening in the periodic and in the RDMFE calculations? If not, why does the periodic DOS exhibit more fine structure?}
   }\label{fig:dos_comp_8atom_16atom}
    \end{figure}
% \XR{I don't quite understand Matthias' comment on the DOS at -10 eV: "Explain why this is not zero?!"} \PR{He seems to think that between the leftmost peak and the 2nd peak there should be a "band gap" in silicon. But I don't think that's the case. That's they only way I can interpret his comment. I'd say we ignore it!}

When comparing the converged, embedded DOS for the 2 atom unit cell with the periodic PBE and PBE0 DOS shown in Fig.~\ref{fig:dos_comp_2atom}, we observe that the band width and band gap are larger than in PBE and are closer to the PBE0 reference.
  When increasing the unit cell size to 8 and 16 atoms (see Fig. \ref{fig:dos_comp_8atom_16atom}) the
difference between the embedded DOS and the periodic PBE0 DOS reduces systematically. The resulting RDMFE band gaps for different 
unit cell sizes are compared with the PBE and the PBE0 values in Tab.~\ref{band_gaps_PBEh}. With increasing unit cell size,
the band gap increases and approaches the PBE0 value.
%\XR{Here we claim that the DOS is converged with the 16 atom unit cell, which is great. However, in the upper
%panel of Fig. 9 we see that the total energy is still far from converging to the reference value with 32 atom 
%unit cell. The common wisdom is however that the total energy converges faster than DOS with the system size.
%What do you think?}
%\Wael{I agree, maybe we should also mention that. But I an not sure how we should do it...}
 \begin{table*}[htp]
\centering
%\scalebox{0.7}{
\begin{tabular}{c c c| c c c | c} \hline	
\textbf{} &  &&\multicolumn{3}{c|}{ RDMFE@PBE0 }&\\
\textbf{} & PBE &PBE0 & 2 atoms &8 atoms &16 atoms &experiment (at 300K)\cite{Ioffedatabase} \\
\hline
\hline
 band gap [eV] & 0.68 & 1.85 & 1.2 & 1.257 & 1.569& 1.12\\
\hline
\hline
\end{tabular}
%}
 \caption{Comparison between PBE, PBE0 and RDMFE for different unit cell sizes for the indirect band gap of silicon. The experimental value\cite{Ioffedatabase} is shown for reference. }
 \label{band_gaps_PBEh}
\end{table*}

\begin{figure}[htp]
   	    \centering
      \begin{tabular}{ccc}
        \includegraphics[scale=0.37]{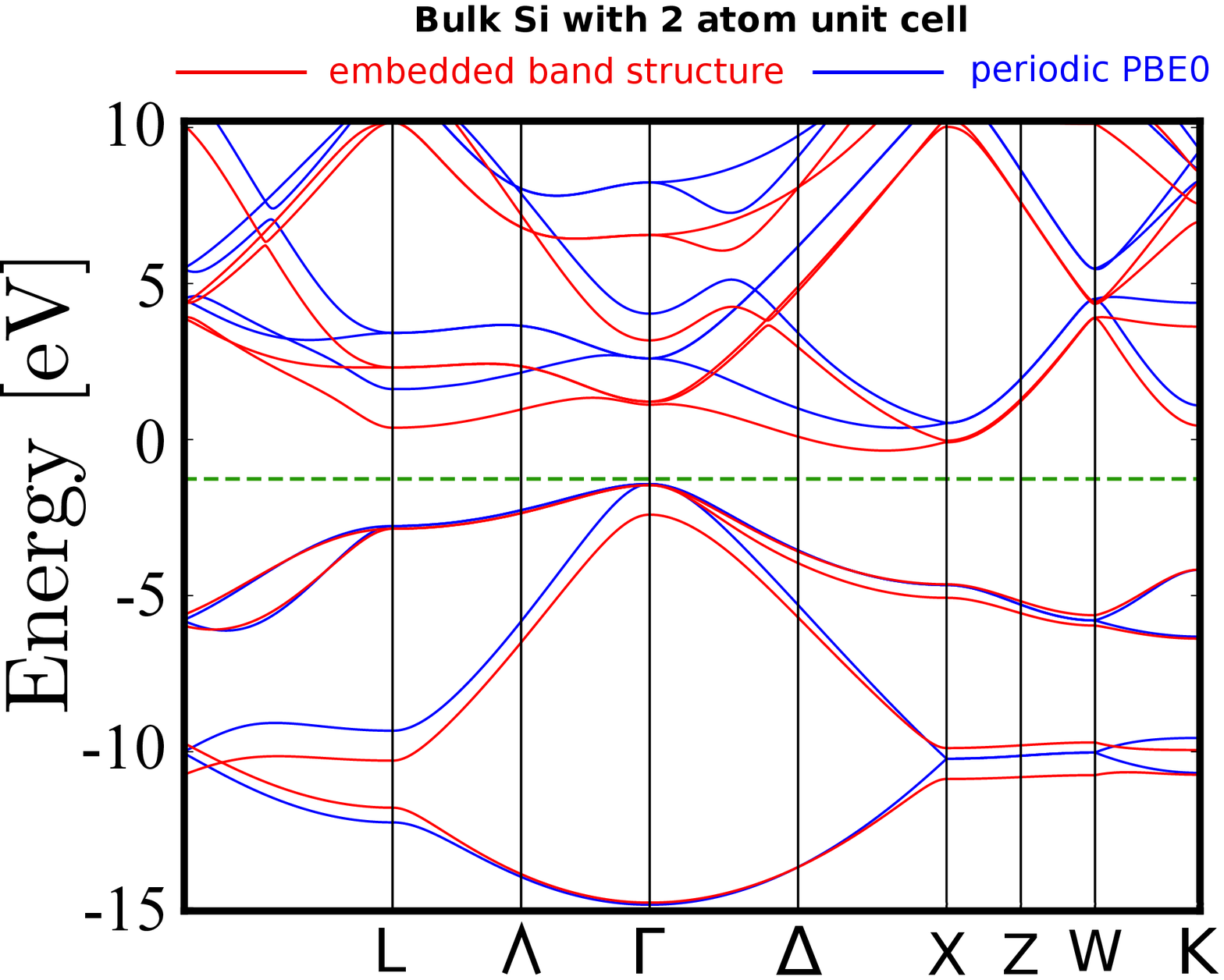}\\
              \includegraphics[scale=0.2]{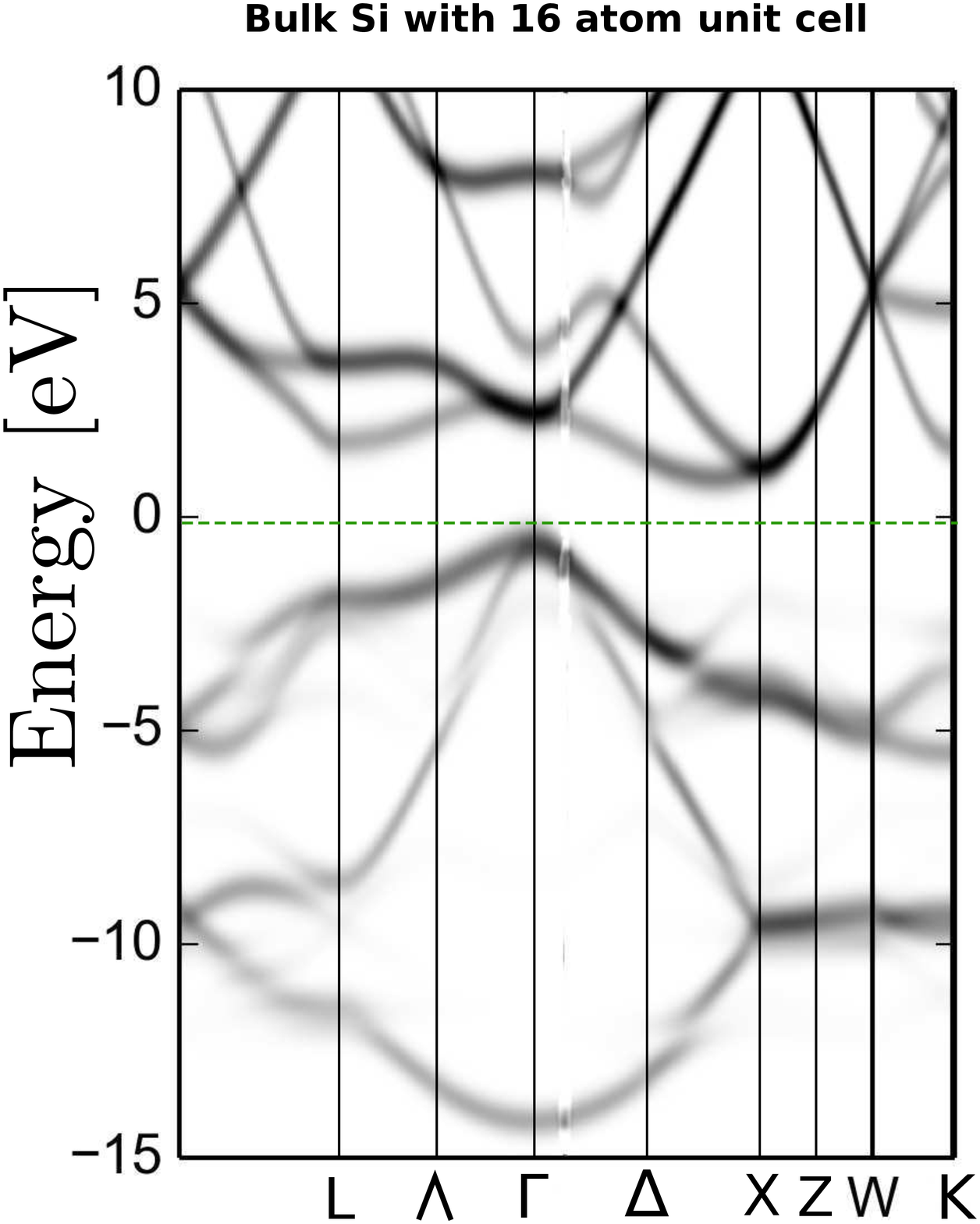}\\
              \includegraphics[scale=0.66]{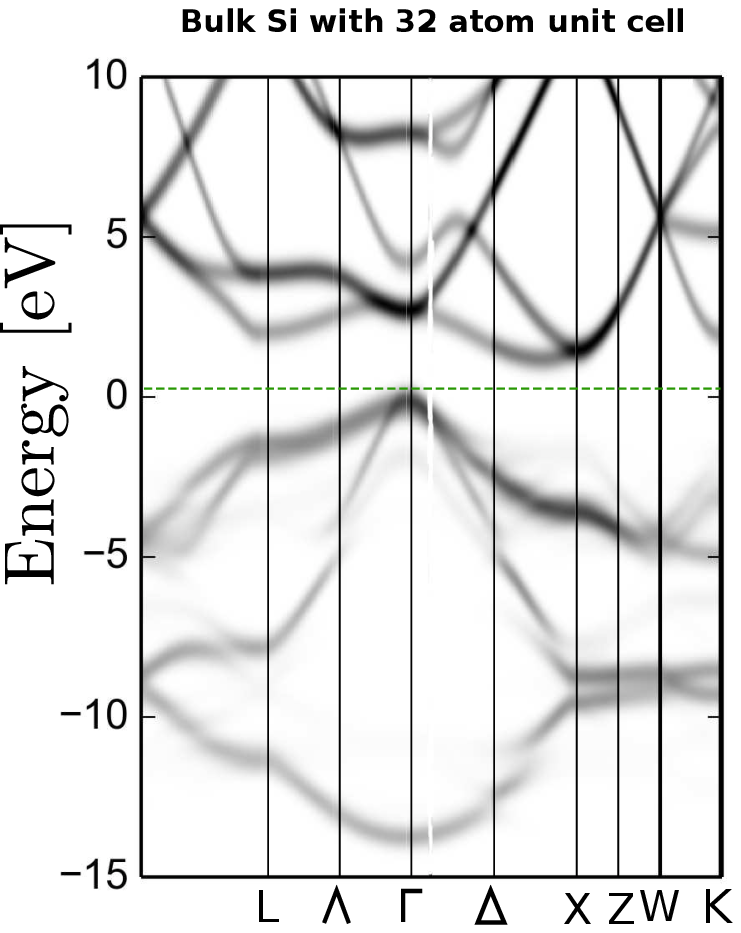}
              \end{tabular}
   \caption{Upper panel: the embedded band structure for bulk Si compared to the periodic PBE0 one. The local self-energy breaks the translation symmetry
            and the degeneracy gets shifted at some hight symmetry $\bfk$-points.
            Middle panel: 
%             \PR{could you increase the line thickness? Some bands are hard to see.} 
            the embedded unfolded band structure for the 16 atom unit cell. The degeneracy shifting gets reduced compared
            to the 2 atoms case. Lower panel: the embedded unfolded band structure for the 32 atom unit cell. Also here the degeneracy is restored in most of the 
            high symmetry points.}
            \label{fig:bandstruct_comp} 
%             \XR{I am still interested to know, for the 8 atom cubic unit cell, if the further dengeracy splitting for 16 atoms still exists?}
    \end{figure}

 %-------------------------------------------------------------------------------------------
Next, we consider the band structure for the 2 atom unit cell shown in the upper panel of Fig.~\ref{fig:bandstruct_comp}. 
We see the same trend as for the DOS: the band gap and the band width approach PBE0 and so do the bands in general. 
However, at some high symmetry points the degeneracy of certain bands is lifted. The origin of this degeneracy lifting is the break of the crystal symmetry that we introduce with the local self-energy. It is a well known artefact and has been discussed extensively in the context of cellular and cluster DMFT \cite{Biroli/etal:2004,Lichtenstein/Katsnelson:2000}. The local self-energy simply does not ``know" about the symmetry of the crystal and can therefore not enforce it. 
The solution to the problem is then obvious: the approximation of the locality of the self-energy needs to be improved. If the self-energy would extend over a larger region    (i.e. supercell) it would acquire more information about the crystal symmetry. 
Then the degeneracy splitting should reduce. 
In the two lower panels of Fig. \ref{fig:bandstruct_comp}, we present an unfolded band structure \cite{unfolding} for the 16 and 32 atom unit cells.  We indeed observe a reduction in the splitting for both the 16 and 32 atom unit cells. However, while the degeneracy is fully restored for some high symmetry points, it is
still broken for others such as the X and Z points. 
 \begin{figure}[htp]
 	    \centering
 \includegraphics[scale=0.3]{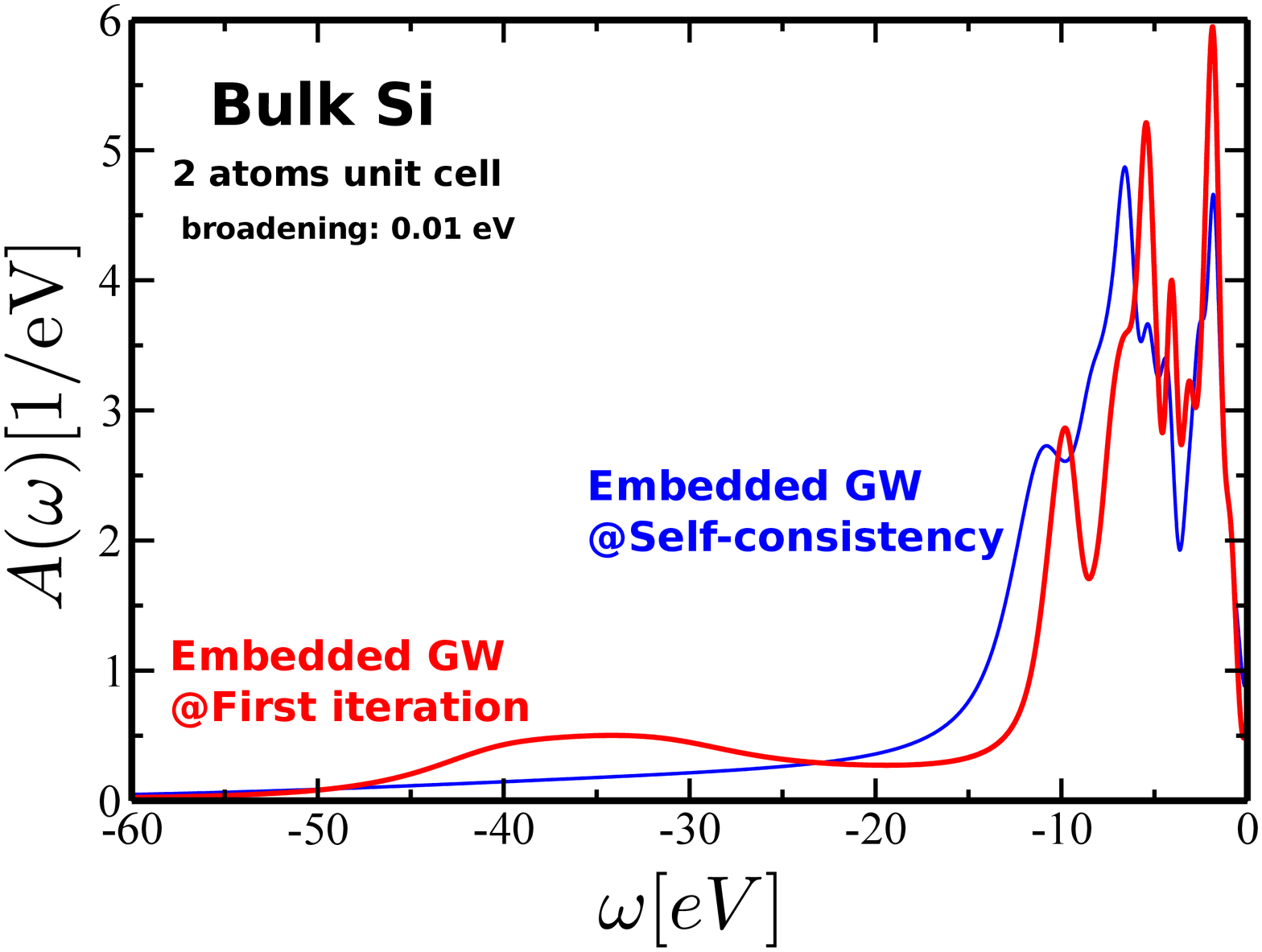}
 \caption{Gaussian broadened (with broadening $\sigma=0.01$ eV) quasiparticle spectrum for the \textit{GW} self-energy at 
  1st iteration (red curve) and at self-consistency (blue curve). Only occupied states are shown.}\label{GW_dos}
\end{figure}
% \XR{It would be interesting to see if GW total-energy flats quickly as the PBE0 one. What happens for helium
% for which you would be able to handle more atoms?}

We will now turn to the $GW$ spectra. The total spectral function for bulk Si with 2 atoms in the unit cell at the 1st iteration is shown in Fig.~\ref{GW_dos}. Since Dyson's equation has been solved once, this spectrum is not equivalent to perturbative $G_0W_0$ spectra and we would expect to see plasmon satellites. The spectrum shows a broad peak between -40 and -30~eV, which has been identified as plasmon satellite. 
\cite{Guzzo/etal:2011,Lischner/etal:2013} Such satellites are completely absent in KS band structures or in $G_0W_0$, because $G_0W_0$ only corrects the KS states and does not yield new states. 
The energy range of the RDMFE satellite agrees well with previous periodic $GW$ calculations \cite{Guzzo/etal:2011,Lischner/etal:2013} and demonstrates that our dynamic, local RDMFE framework can capture non-local phenomena such as plasmon satellites. 
For sc\textit{GW} the converged DOS is also shown in Fig. \ref{GW_dos}. As demonstrated by Holm and von Barth \cite{Holm/vonBarth:1998} for the electron gas, 
full self-consistency in $G$ and $W$ leads to a deterioration in the \textit{GW} spectral function due to the neglect of vertex corrections. Thus, the fact, that the plasmon satellite disappears at self-consistency
is not surprising.

We obtain a band gap of $\sim$0.9 eV for the two atom unit cell, which is close to the experimental value of $\sim$1.12 eV \cite{Kasap/Capper:2006}. 
This comparison together with the one between the indirect band gap from our calculation and experiment\cite{Ioffedatabase} are presented in 
Tab. \ref{band_gaps_GW}.
% \XR{Here we need to explain a bit why both $G_0W_0$ and sc$GW$ don't yield the plasmon satellite, but the 1st
% iteration $GW$ does. }
% \PR{Can you say a bit more here? Maybe do a mini table with the direct and indirect gaps in scGW and experiment. Then you can also include the RDMFE PBE0 values for 2, 8 and 16 atom cells and discuss all numbers a little bit in this paragraph. } \XR{I see that there is no RDMFE PBE0 values yet.}
%It is also worth mentioning that to obtain the quasi particle spectrum analytical continuation was used.\\  
% \begin{center}
\begin{table*}[htp]
\centering
%\scalebox{0.7}{
\begin{tabular}{  c c c c c} \hline
\textbf{band gap } & RDMFE@sc\textit{GW} &periodic sc\textit{GW}\cite{Kutepov:2009}& $QP$sc\textit{GW}\cite{Kotani/etal:2007}& experiment (at 300K) \\
\hline
\hline
 direct ($\Gamma_{15c}$) [eV]                                & 3.7 &|&3.47& 3.4 \cite{Ioffedatabase}\\
 indirect ($E_{g}$) [eV]                              &0.9 &1.55&1.25& 1.12 \cite{Kasap/Capper:2006}\\ %\hline\hline
\hline
\hline
\end{tabular}
%}
 \caption{Direct and indirect band gaps as calculated from RDMFE for the \textit{GW} self-energy. Comparison is made with the periodic sc\textit{GW} work of 
 Kutepov \textit{et al.}\cite{Kutepov:2009} and the quasi particle self-consistent \textit{GW} calculation of Kotani \textit{et al.}\cite{Kotani/etal:2007}
 and experiment.}
 \label{band_gaps_GW}
\end{table*}
 %-------------------------------------------------------------------------------------------

 \subsection{Embedded total energies:} 

For sc$GW$ we currently do not have a periodic reference to compare to, as alluded to before. We can, however, construct another test case and benchmark against our sc$GW$ 
implementation for finite systems\cite{Caruso/etal:2012,Caruso/etal:2013_tech}, where the total energy was computed from the Galitskii-Migdal formula \cite{Galitskii/Migdal:1958}.
We achieve this by considering the molecular limit of a unit cell, i.e., the limit of an isolated unit 
cell with a lattice constant of $\sim$20~\AA{}. The benchmark results for He, H$_{2}$ and Na$_{2}$ are presented in Tab.~\ref{E_GW_benchmark}, 
which shows the XC components that enter the total energy as given by Eq. (\ref{gw_xc_contribution_GM}).
$\Sigma^{\text{sc}GW}_{\rm XC}$ is the molecular sc$GW$ XC self-energy and $G^{scGW}(i\omega)$ the corresponding Green's function at convergence. 
$\Sigma^{\rm loc}_{\rm XC}$ is local XC self-energy and $G^{\rm emb}(i\omega)$ the embedded Green's function both obtained at convergence of the RDMFE cycle.
% The local embedded exchange and correlation self-energies $\Sigma^{\rm loc}_{xc}$ and the sc$GW$ ones $\Sigma^{\text{sc}GW}_{xc}$
% are considered at convergence. \PR{I find this notational explanation extremely tedious! Aren't all the terms, at least for the embedded quantities, defined already in the theory section? Why repeat them? Or even worse, why use slightly different symbols?}
Table~\ref{E_GW_benchmark} illustrates that the components entering the embedded total energy agree almost to the meV level with the corresponding components
from the finite systems sc\textit{GW} calculation, which demonstrates the reliability and robustness of our implementation.

\begin{figure}[htp]
   	    \centering
      \begin{tabular}{cc}
        \includegraphics[scale=0.3]{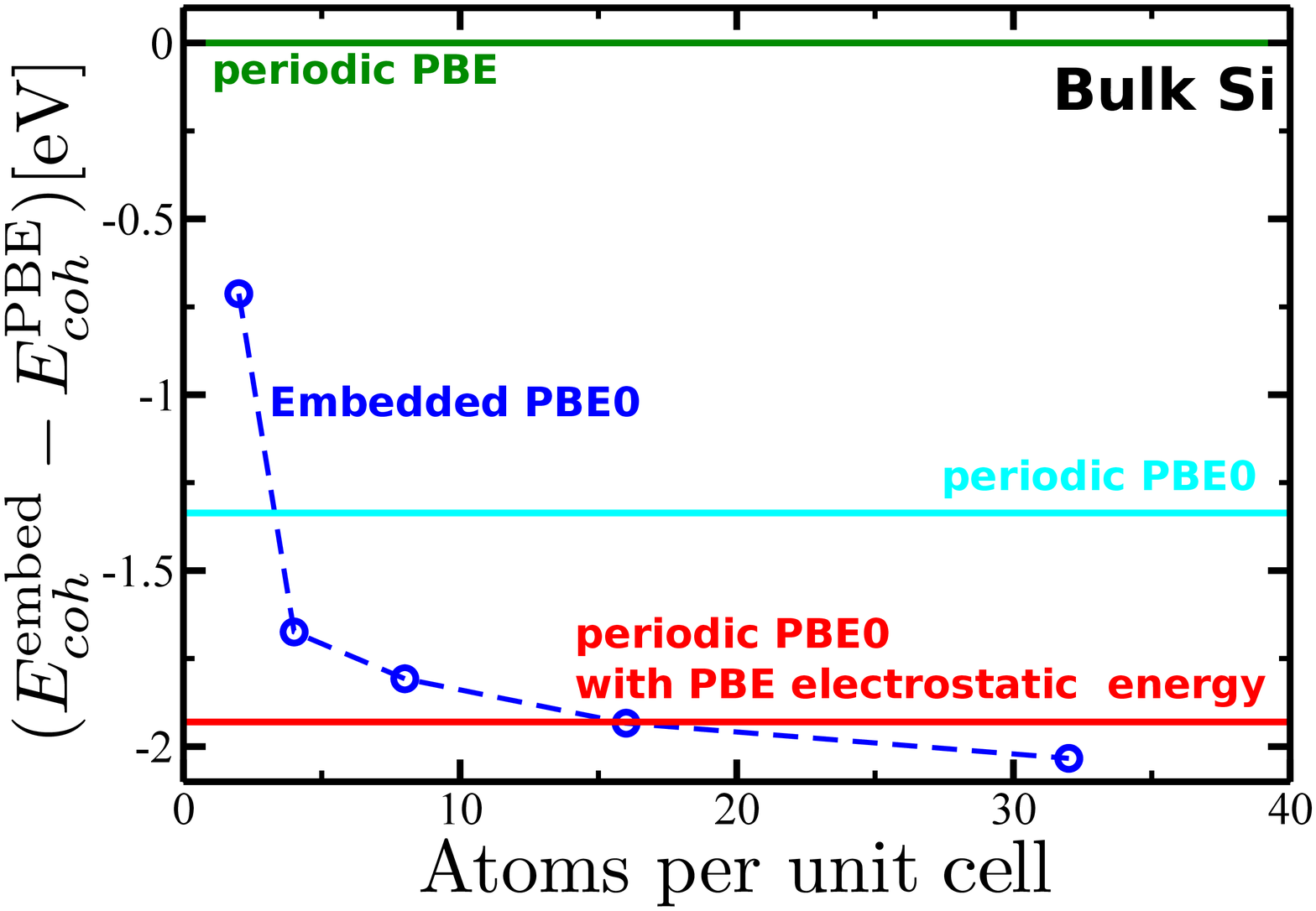}
      \\ \includegraphics[scale=0.3]{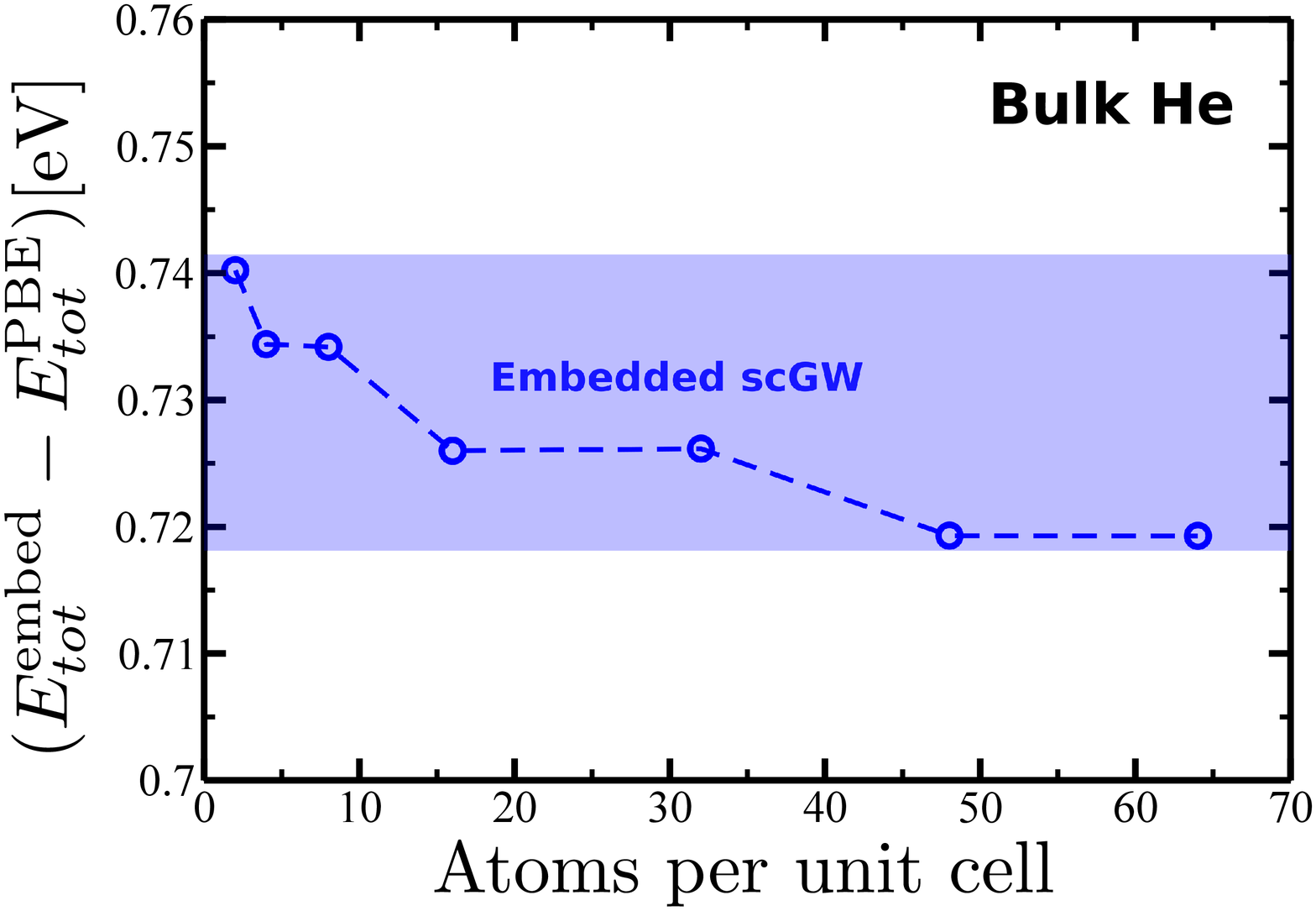}
      \end{tabular}
   \caption{Upper panel: the embedded PBE0 cohesive energy for bulk Si for increasing unit cell size.
            Lower panel: the embedded \textit{GW} total energy for bulk He for increasing unit cell size, where the blue region indicates a change in the range of $\sim 20$ meV. Both curves are referenced to the PBE total energy.}\label{fig:total_energy_comp}
    \end{figure}

\begin{figure}[htp]
       	    \centering
           \begin{tabular}{cc}
        \includegraphics[scale=0.3]{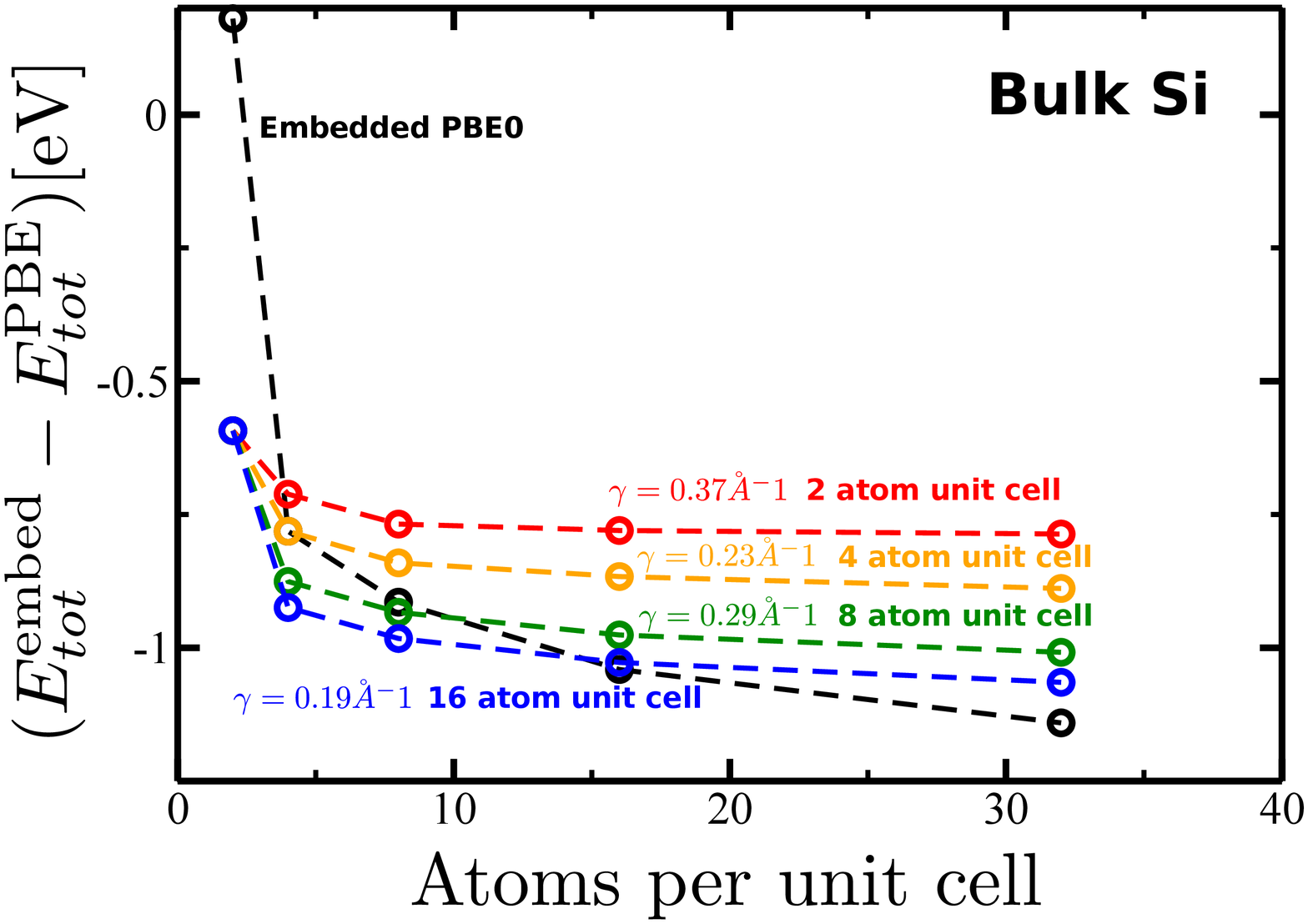}
      \\ \includegraphics[scale=0.3]{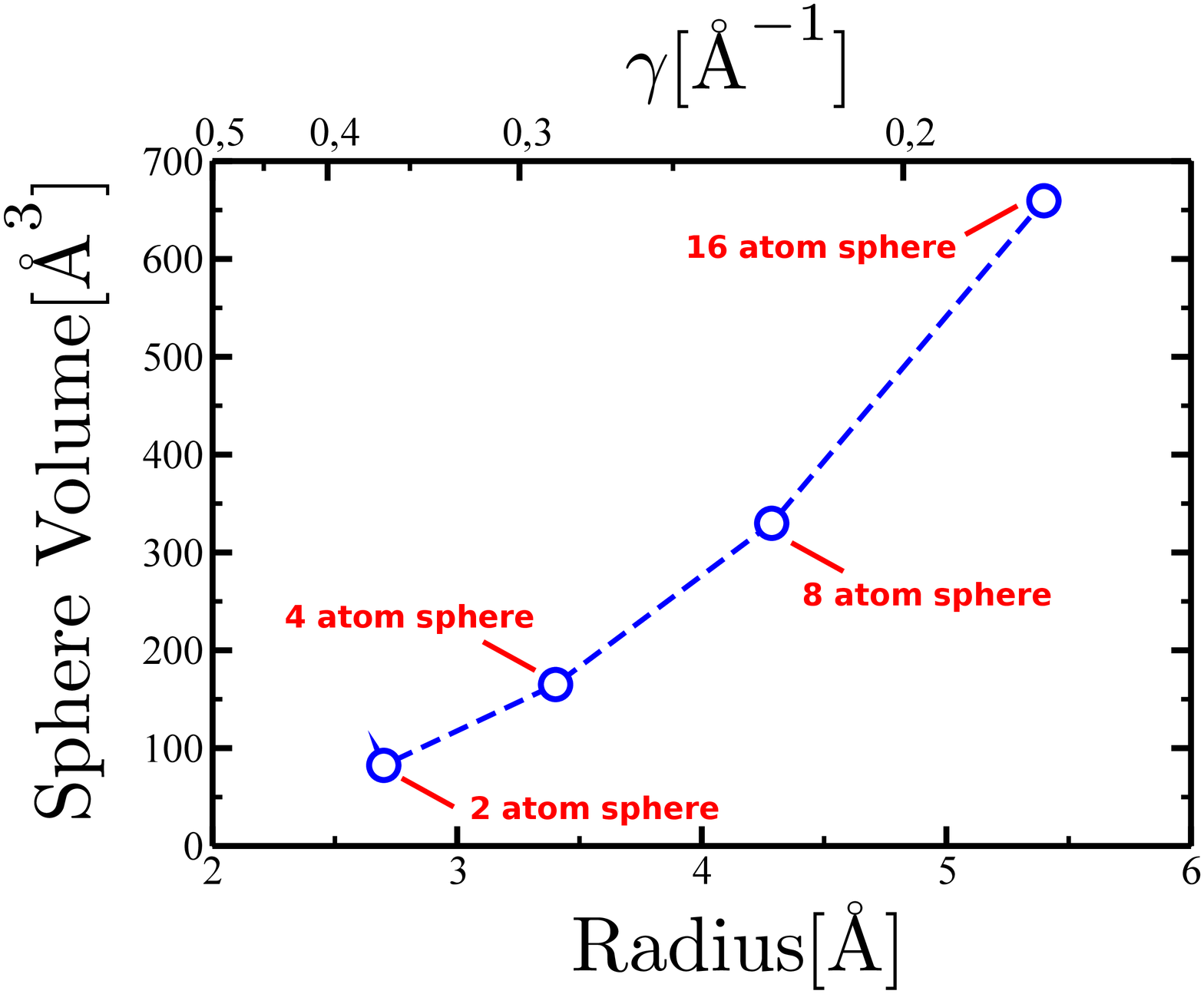}
      \end{tabular}
   \caption{Upper panel: the embedded HSE total energy for bulk Si for increasing unit cell size and different screening parameter $\gamma$.  The range separated total energy converges faster than the PBE0 one (black curve).
	    All the embedded HSE total energies are shifted to the 2 atom unit cell value of the embedded PBE0 curve to allow better comparison. 
            Lower panel: The change of the sphere radius and screening parameter $\gamma$ (on a reciprocal scale) with the volume of the sphere surrounding the 2, 4, 8 and 16 atom unit cell.\label{fig:total_energy_comp_hse}
%             \XR{What you showed in the lower pannel is the sphere volume as a function of the sphere radius,
% and not the screening parameter.} \PR{ 
% In the lower panel, I would use the upper x-axis of the figure and plot omega on it. So you have the radius on the lower x-axis and on the upper one you see how $\omega$ changes with radius. }
% \XR{I am wondering, why none of the embedded HSE calculations with all different screnning parameters goes to
% the periodic HSE limit? Does this indicates there is a flaw in our total energy calculation?}
}
           \end{figure}

% \begin{center}
\begin{table*}[htp]
\centering
%\scalebox{0.7}{
\begin{tabular}{ c c c c} \hline
\textbf{Term} & \textbf{He} & \textbf{H$_{2}$}& \textbf{Na$_{2}$}\\
\hline
\hline
\\
%  $ H n^{\rm emb} $                                       &-30.971399  & -20.082893   & -4681.324289 \\ %\hline\hline
%  $H n^{scGW}$                                          & -30.976581 & -20.084208      & -4681.324755   \\ %\hline\hline
% \\
%  $v^{KS}_{\rm XC} n^{\rm emb}$                               & -36.283999 & -24.105587   & -1004.447234 \\ %\hline\hline
%  $v^{KS}_{\rm XC} n^{scGW}$                                & -36.286584 & -24.106258     & -1004.447806 \\ %\hline\hline
% \\
%  $v^{KS}_{H} n^{\rm emb}$                                & 54.994712  &  35.697015    & 4847.058775\\ %\hline\hline
%  $v^{KS}_{H} n^{scGW}$                                 & 54.996847  &  35.697851      & 4847.073781 \\ %\hline\hline
% \\
%  $v^{\rm emb}_{H} n^{\rm emb}$                             & 54.835947  &  35.690403    &  4849.984330 \\ %\hline\hline
%  $v^{scGW}_{\rm H} n^{ scGW}$                              & 54.843715  &  35.692794     &  4849.995830  \\ %\hline\hline
% \\
%  $\Sigma^{\rm loc}_{\rm X} n^{\rm emb}$                        &-27.173906  & -17.503642  & -760.630196\\ %\hline%\hline
%  $\Sigma^{scGW}_{\rm X} n^{ scGW}$                          &-27.181325  & -17.505250     & -760.632183 \\ %\hline\hline
% \\
%  $\Sigma^{\rm loc}_{\rm C} n^{\rm emb}$                        &-1.744578   & -2.3053908    & -2.457104 \\ %\\ \hline %\\ %\hline\hline
%  $\Sigma^{scGW}_{\rm C} n^{ scGW}$                          &-1.737910   & -2.304200      & -2.455707 \\ %\hline%\hline
% \\
% \hline
% \hline
% \\
 $E^{GW}_{\rm X}[G^{\rm emb}]=\frac{1}{2}\int \frac{d\omega}{2\pi i}\Sigma^{\rm loc}_{\rm X}G^{\rm emb}(i\omega)e^{-i\omega 0^{+}}$                    &-27.173906  & -17.503642  & -760.630196\\ %\hline%\hline
 $E^{GW}_{\rm X}[G^{scGW}]=\frac{1}{2}\int \frac{d\omega}{2\pi i}\Sigma^{scGW}_{\rm X}G^{scGW}(i\omega)e^{-i\omega 0^{+}}$                          &-27.181325  & -17.505250     & -760.632183 \\ %\hline\hline
\\
 $E^{GW}_{\rm C}[G^{\rm emb}]=\frac{1}{2}\int \frac{d\omega}{2\pi i}\Sigma^{\rm loc}_{\rm C}(i\omega)G^{\rm emb}(i\omega)e^{-i\omega 0^{+}}$                     &-1.744578   & -2.3053908    & -2.457104 \\ %\\ \hline %\\ %\hline\hline
 $E^{GW}_{\rm C}[G^{scGW}]=\frac{1}{2}\int \frac{d\omega}{2\pi i}\Sigma^{scGW}_{\rm C}(i\omega)G^{scGW}(i\omega)e^{-i\omega 0^{+}}$                           &-1.737910   & -2.304200      & -2.455707 \\ %\hline%\hline
\\
\hline
\hline
\\
%  Total energy $E^{\rm emb}_{\rm GM}$                           &-78.759362  & -32.067732  & -8808.470968 \\ %\hline\hline
%  Total energy $E^{scGW}_{\rm GM}$                            &-78.759212  & -32.068074     & -8808.489963  %\hline\hline
\end{tabular}
%}
 \caption{Exchange (X) and correlation (C) components of the RDMFE@sc\textit{GW} total energy, $E^{GW}_{\rm XC}[G^{\rm emb}]$, as given by Eq. (\ref{gw_xc_contribution_GM}) in the limit
 of an isolated unit cell (large lattice constants): 
 a benchmark against the standard sc\textit{GW} calculation for finite systems\cite{Caruso/etal:2012}, labeled by $E^{GW}_{\rm XC}[G^{scGW}]$. All energies are in eV. 
% \PR{Did you update this table, when we updated the total energy scheme last month? In the caption you still mention the GM scheme, but we don't really refer to GM anymore in the total energy section.}
% \Wael{This table was not updated, since it includes the comparison with the total energy formula as it was implemented in FHI-aims for \textit{scGW} (i.e. the GM formula).
% While, for our total energy we only need
% the XC parts and the kinetic parts, I think this comparison is still a very important check for our Green functions and the resulting density matrices.}
}
 \label{E_GW_benchmark}
\end{table*}
% \XR{One should tell the reader what the different terms mean, and where the remaining difference comes from.}
% \end{center}
%-------------------------------------------------------------------------------------------
We then investigated the convergence of the total energy with respect to the increase of the unit cell size for RDMFE PBE0 and sc\textit{GW}. For embedded PBE0, we performed calculations for bulk Si up to 32 atoms in the unit cell, 
% \PR{If you have this 32 atom calculation, why don't you show that also in the DOS section? It would be great, if you could unfold the 32 atom band structure instead of the 16 atom band structure!}
 whereas for \textit{GW} we considered bulk He in the fcc structure up to 64 atom unit cells.
To reach larger systems, a full parallelization of our implementation would be required.
The upper panel of Fig.~\ref{fig:total_energy_comp} shows the comparison of our embedded PBE0 cohesive energy  with
the periodic PBE and the periodic PBE0 energy. 
We also include a third reference in which we added the kinetic and XC energy of a PBE0 calculation to the PBE energy, which most closely resembles our RDMFE approximation. 
We see that with increasing unit cell size the embedded cohesive energy approaches the periodic PBE0 value, but then dips below. 
This is not surprising since our embedded cohesive energy does not account for changes in the electrostatic energy. 
Instead, the RDMFE curve approaches the PBE0 reference value from which the electrostatic change has been removed. 
However, the convergence to the periodic limit is relatively slow. 
This can be related to the long range nature of the HF exact-exchange as we will show later on using a range separated self-energy (see the discussion of Fig.~\ref{fig:total_energy_comp_hse}).  For the \textit{GW} self-energy, however, the total energy seems to converge much faster and only changes in a very small range. 
This  is shown in the lower panel of Fig. \ref{fig:total_energy_comp}.

% \XR{Plotted in the upper panel are the HSE total energy with different screening length, correct? The discussion
% in the text isn't very clear. I can see that by changing the screening length, the total energy gets shifted. But
% don't see, e.g., that ``2 atom sphere" converges faster than the ``16 atom sphere".   }
To better visualise the interplay between locality and unit cell size in our scheme we also present results of the HSE range-separated exact-exchange self-energy. 
We vary the range separation parameter to model different degrees of locality, but keep the percentage $\alpha$ of exact exchanged fixed. 
We then translate the range separation parameter into a radius $R_{\rm sphere}$ in real-space using the relation $\gamma=R^{-1}_{\rm sphere}$ and determine the number of atoms that fit inside. 
We have considered range-separation parameters $\gamma$ that correspond to spheres enclosing  2, 4, 8 and 16-atom unit cells. 
For each $\gamma$, the resulting embedded total energy is plotted in Fig.~\ref{fig:total_energy_comp_hse} as a function of the size of the unit cell (upper panel). 
The lower panel shows the volume of the surrounding sphere for the different $\gamma$ parameters and for the different unit cell sizes. 
We indeed observe that the total energy converges faster with unit cell size, the shorter the range of the non-locality in the HSE self-energy. 
This proofs that RDMFE becomes a viable option for self-energies, whose range only encompasses a few nearest atoms. 
In that sense, PBE0 had been the toughest test, because its range is infinite.

% \XR{It seems the embedded HSE total energy for different screening parameters are converged with 32 atom unit cell. 
% I am wondering they converge to the periodic HSE value?} \PR{Yes, can you please include the periodic reference values in Fig.~\ref{fig:total_energy_comp_hse}, too?}

\begin{figure}[htp]
       	    \centering
      \begin{tabular}{cc}
        \includegraphics[scale=0.3]{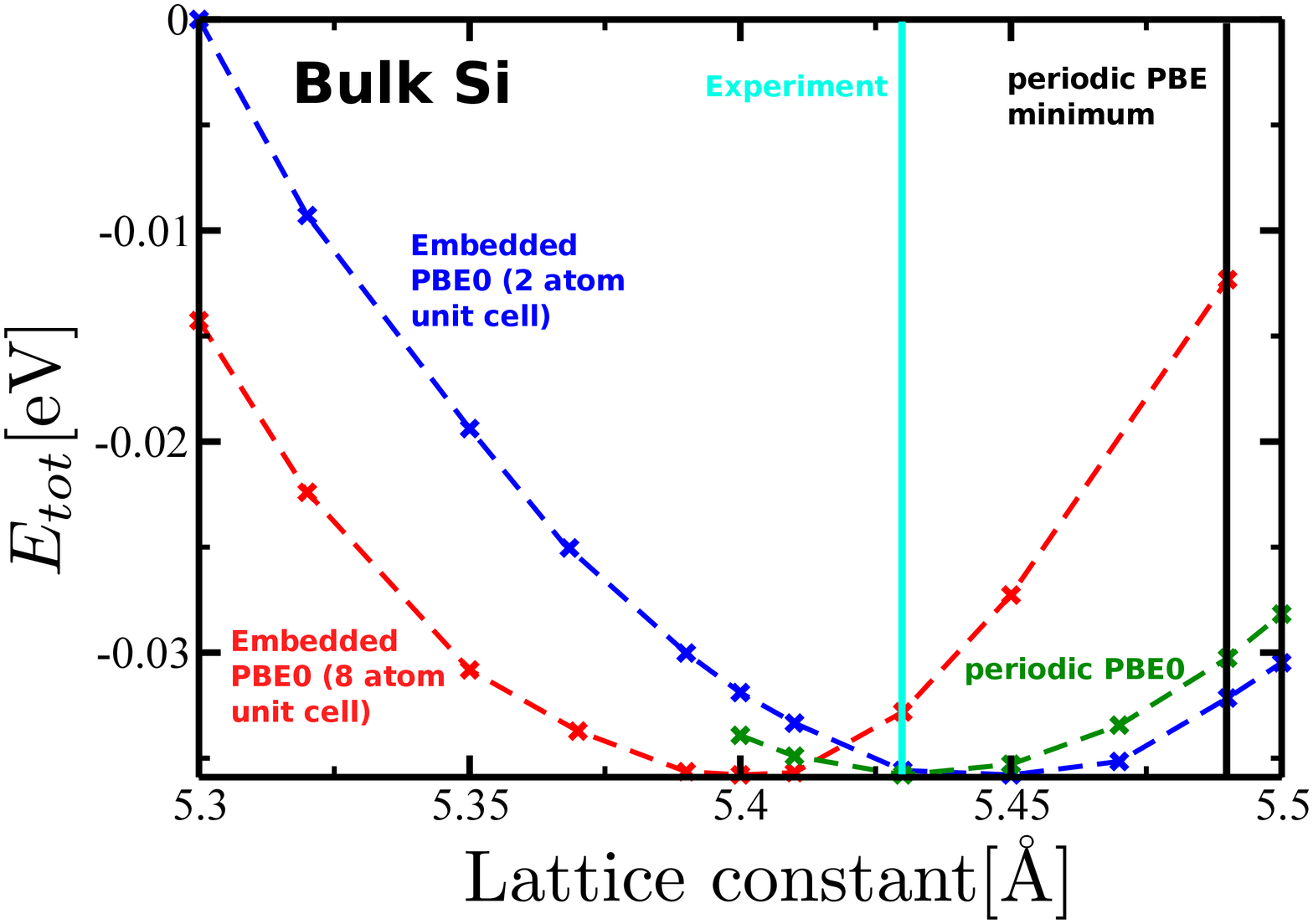}
      \\ \includegraphics[scale=0.3]{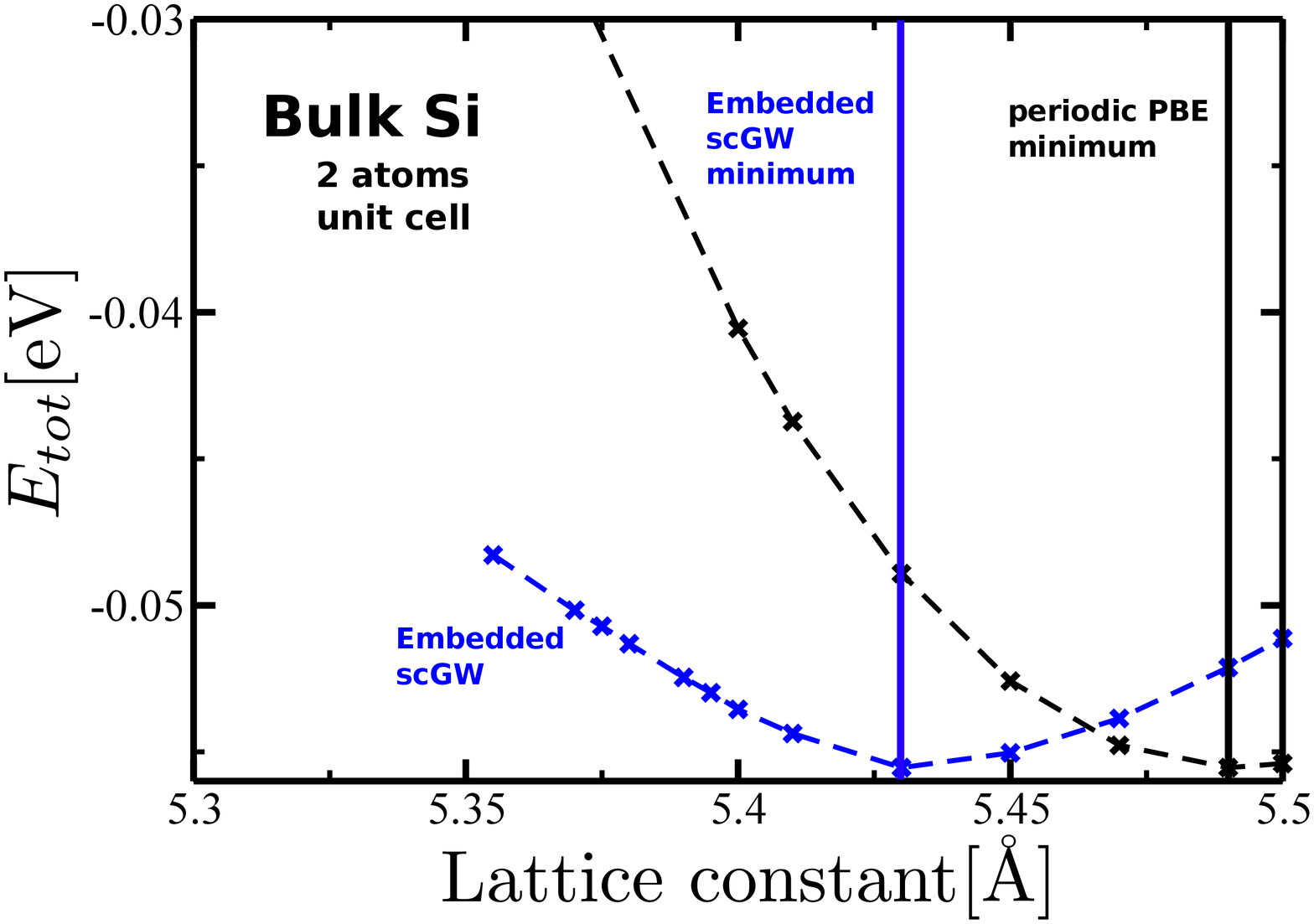}
      \end{tabular}
   \caption{The embedded total energy for bulk Si for different lattice constants: RDMFE@PBE0 is shown in the upper and RDMFE@$GW$ in the lower panel.
%    \PR{Could you please use a consistent color code? In the upper panel you use green for PBE0, but in the lower panel it denotes PBE. Could you also mark the experimental equilibrium lattice constant in the upper panel?}
   }\label{fig:bind_energy_comp}
    \end{figure}

Finally, we briefly address cohesive properties. In Fig.~\ref{fig:bind_energy_comp} we show the total energy of bulk Si as a function of the lattice constant for RDMFE@PBE0, scGW and for periodic PBE0. For RDMFE@PBE0 (upper panel) our calculations for a 2 atom unit cell already give a minimum that is below the PBE one and very close to the experimental value of 5.43\AA \cite{Hom:1975} (see Tab.~\ref{lattice_constants}). However, for an 8 atom unit cell the lattice constant reduces slightly. For RDMFE@$GW$, the minimum for the 2 atom unit cell (lower panel) also lies below the PBE value and already agrees fortuitously well with the experimental value.  Our RDMFE@$GW$ lattice constant is slightly larger than that reported by a recent periodic self-consistent $\textit{GW}$ calculation\cite{Kutepov:2009}. We also performed a Birch-Murnaghan \cite{Birch:1947} fit of the total energy curves to extract the bulk moduli for RDMFE@PBE0 and RDMFE@$GW$. The resulting values are reported in Tab.~\ref{lattice_constants}.

\begin{table*}[htp]
\centering
  \begin{tabular}{c||c|c|c||c|c | c | c} 
%       \hline	
	  \textbf{} &\multicolumn{2}{c}{RDMFE@PBE0} & periodic PBE0 & RDMFE@sc\textit{GW} & periodic sc\textit{GW}\cite{Kutepov/etal:2012} & experiment \\
      \hline
      \hline
	 unit cell size &2 atoms & 8 atoms & 2 atoms & 2 atoms   & | & | \\
%  Indirect                               &0.9 & 1.12 \cite{Kasap/Capper:2006}\\ %\hline\hline
%       \hline
%       \hline
      	  lattice constant [\AA] &5.45 & 5.4 & 5.43 & 5.43 & 5.39 & 5.43\cite{Hom:1975}\\
%  Indirect                               &0.9 & 1.12 \cite{Kasap/Capper:2006}\\ %\hline\hline
      	  bulk modulus $B_{0}$ [GPa] &95.14 & 129.07 & 84 & 80.37& 100.7 & 99\cite{Rodriguez/etal:1985}\\
      \hline
      \hline
  \end{tabular}
%}
 \caption{Bulk Si RDMFE equilibrium lattice constant and bulk moduli $B_{0}$ for the embedded PBE0 and \textit{GW} self-energies. Comparison is
 made with periodic PBE0 performed with FHI-aims, the periodic sc\textit{GW} work of Kutepov \textit{et al.}\cite{Kutepov/etal:2012} and experiment.\cite{Hom:1975,Rodriguez/etal:1985}}
 \label{lattice_constants}
\end{table*}
 %-------------------------------------------------------------------------------------------

\section{\label{sec:discussion}Discussion}
We have presented an embedding scheme for periodic systems that builds on DMFT. 
In our approach, the electron interacting across periodically repeated unit cells is mapped  onto an on-site problem, in which the electrons only interact directly in one unit cell, but are dynamically coupled to a periodic bath of electrons. The coupling between the embedded system and the surrounding is constructed naturally by means of Green's functions. Due to its dynamic nature the bath can exchange electrons with the embedded region.
Our embedding scheme is most suitable for systems with 
periodic boundary conditions, as the translational symmetry is automatically preserved.
 Furthermore, we transfer the non-locality of a self-energy into a frequency dependence, a concept that has previously been explored by Gatti {\it et al.} \cite{Gatti/etal:2007_nonlocality} 
and in the spectral density-functional theory of Kotliar {\it et al.} \cite{Savrasov/Kotliar:2004}
We note that the only approximation introduced in our scheme is that the non-local XC 
coupling between neighboring unit cells (or computational supercells) is included only at the KS GGA level, and 
neglected in the more advanced (here hybrid functional or $GW$) treatment. In other words, 
the self-energy correction to the GGA XC potential is $\bfk$-independent, an intrinsic feature of
DMFT.

We now compare our scheme to other embedding schemes.  For the hybrid QM:MM approach a clear separation between the embedded region and the surrounding and the treatment of  the boundary atoms  is not always obvious\cite{Hu/eta_erratum:2007,Scheffler/etal:1985}.
%Moreover, the fact, that the surrounding is not treated quantum mechanically leads to a considerable loss of accuracy.
For systems, in which classical electrostatics dominate such 
as ionic or molecular solids, the separation between ions and molecules is natural. However,  for 
covalently bonded systems it becomes more difficult to define the QM:MM partitioning.
Thus, typically covalent bonds are cut at the QM:MM boundary, which produces dangling bonds that need to be saturated. 
A multitude of models with different levels of accuracies
have been developed to tackle these issues.
One example is the ChemShell framework \cite{Sherwood20031,QUA:QUA20032} that supports Hartree-Fock and hybrid functionals in the embedded region and that has recently been coupled to FHI-aims \cite{Berger/etal:2014}.

Another popular approach is "Our own \textit{N}-layer integrated molecular orbital molecular mechanics" (ONIOM)  by Morokuma and coworkers. \cite{Maseras/Morokuma:1995,Chung/etal:2012}  ONIOM is a so-called 
extrapolative (or subtractive) scheme in which the total energy of the whole system is given by
   		      \begin{align}           
   		        E_{\rm ONIOM} = E_{\rm RL} - E_{\rm ML}+E_{\rm MH},
   		      \end{align}
where the RL refers to the real (or full) system at the lower level, ML refers to the model (or embedded) system at the lower level and MH labels the model system for the higher level theory. In constrat to the additive
QM:MM scheme, ONIOM does not need an additional coupling Hamiltonian to describe the QM/MM interation.
When a QM/MM boundary cuts through a covalent bond, link atoms (mostly hydrogen atoms) are added to
cap the unsaturated QM boundary for the model calculations.
Even if it is common to use MM methods for describing the surroundings,  the ONIOM scheme was recently
extended to deal with two-layer two-QM embedding, ONIOM(QM1:QM2), where
HF was used for the surroudings and the embedded model region is described by MP2 or B3LYP. The QM1/QM2
interactions, including electrostatic interaction, mutual polarization, and charge transfer, are 
described at the lower QM level.

Our RDMFE scheme is distinctly different from the ONIOM(QM1:QM2) approach. First, the
RDMFE scheme is formulated in terms of Green's functions, whereas ONIOM is based on a partition of total energies.
As such, spectral properties come out naturally from RDMFE, while the evaluation of total energies is
more involved, as discussed in Sec.~\ref{sec:total_energy}. The opposite is true for the   
ONIOM(QM1:QM2) scheme. Second, within RDMFE, the effect of the environment is encoded in the bath
Green's function that describes an electron reservoir with which the embedded cluster can exchange electrons freely. 
In other words, the electronic states in the embedded system are not forced to localize within the cluster, 
but are allowed to delocalize into the surrounding system. Thus, dangling bonds pose no conceptual problem and
boundary effects are not significant since they diminish quickly as the size of the cluster increases. In contrast,
in ONIOM(QM1:QM2), like in most other embedding schemes in computational chemistry, link atoms are needed to 
saturate the dangling bonds when chemical bonds are broken. Therefore ONIOM(QM1:QM2) is most appropriate for
describing systems with localized electrons, whereas RDMFE has no problem in dealing with delocalized electrons,
especially metallic systems. Third, RDMFE, as is formulated right now, is only applicable to periodic systems 
that are relevant to solid state physics, while ONIOM(QM1:QM2) is most suitable for describing molecules 
and clusters that are of interest to chemical and biological applications.

Addressing the problem of CO adsorption on Cu(111) \cite{Feibelman:2001},
Hu, Reuter and Scheffler \cite{Hu/Reuter/Scheffler:2007,Hu/eta_erratum:2007} developed a cluster extrapolation scheme that is based on performing a cheap (LDA/GGA)
calculation for the periodic system then correcting the resulting total energy by $\Delta E_{\rm XC}=E^{\rm cluster}_{\rm XC}[{\rm LDA/GGA}]-E^{\rm cluster}_{\rm XC}[{\rm ``better"}]$,
where $E^{\rm cluster}_{\rm XC}[{\rm LDA/GGA}]$ and $E^{\rm cluster}_{\rm XC}[{\rm ``better"}]$ are the cluster XC-energy parts of a cluster calculation with the cheaper (LDA/GGA) and the "better" theory
respectively while the cluster itself has been cut out from the periodic system. Increasing the cluster size, they could then show that
the correction $\Delta E_{\rm XC}$ converges for relatively small cluster sizes ($\sim$16 atoms) and thus much faster than $E^{\rm cluster}_{\rm XC}["better"]$ alone. This cluster extrapolation concept is similar to that of
ONIOM(QM1:QM2) described above, but link atoms were not used for the cluster calculations.

Whitten and coworkers also approached molecular adsorbates on metal surfaces.\cite{Whitten/Pakkanen:1980}  They
developed an embedding scheme that builds on identifying a localized subspace that has maximal exchange overlap
with the valence orbitals of the atoms within and bordering the adsorbate. The localized subspace
is then solved using the CI method, for a fixed Coulomb and exchange potential constructed from the localized orbitals. However, the approach mimics
the real periodic system using a large cluster of atoms, which fails in describing the system accurately. 
Additionally, no systematic cluster extrapolation has been studied.
In a similar spirit,  Huang and Carter developed a density-functional-based embedding scheme. \cite{Huang/Carter:2011} The scheme relies on the fact that the density is additive, i.e., that the total electron density can be partitioned into the density of the embedded region
and the density of the embedding surrounding. Proceeding as such, allows the definition of 
an embedding density-potential, that is a functional of the total and the embedded
density. Adding this potential to the embedded Hamiltonian and solving the resulting KS Schr\"odinger
equation self-consistently leads to the desired embedded density. For the embedded region, correlated wave function methods 
are typically used, while for the embedding potential the optimized effective potential method or kinetic energy density functionals
are employed.
% \PR{I guess at some point the embedded region is then treated with a better method, no?!} 
% However, the fact that the embedding potential 
% is a functional of two independent densities requires delicate approximations that are numerically not always stable
% for self-consistency. 
% \XR{I am not aware of this issue. Patrick, do you know a reference that demonstrates this
% problem?} \PR{No. I didn't put this in. Wael, did you? As far as I remember, the actual problem with this approach is that a kinetic energy functional is needed that is given in terms of the density (and not the Kohn-Sham orbitals). Thus, the problem is similar to the ones faced in orbital free DFT. Wael, please check the papers and then change the statement.}
Moreover, due to the static nature of the embedding potential no dynamical methods can be used to describe the embedded region, which limits the applicability to ground state properties.

For point defects in semiconductors, Scheffler \textit{et al.} \cite{Scheffler/etal:1985} devised a self-consistent 
Green's function method to compute the change in density induced by the presence of the defect. They considered this change as being
a perturbation to the perfect crystal and solved the resulting Dyson equation self-consistently. Using the fact that defects are well localized in
real space, they correct the Hellmann-Feynman force of the perfect crystal -- calculated with force fields -- by a contribution containing the change in density due to the defect -- calculated with KS-DFA. 
They showed that the resulting Hellmann-Feynman force is comparable in accuracy to a full DFA calculation.

Finally, it is also worth mentioning, that we perform fully self-consistent \textit{GW} calculations \cite{Ren/etal:2012,Caruso/etal:2013} in our scheme, which is conceptually different from the so called quasiparticle self-consistent \textit{GW} concept\cite{Schilfgaarde/Kotani/Faleev:2006,Kotani/etal:2007} (QPsc\textit{GW}). In QPsc\textit{GW} a series of  $G_{0}W_{0}$ calculations is performed. In each iteration the ``best'' $G_0$ is determined that most closely resembles the $GW$ Green's function of the current cycle. In practice a static, non-local potential is constructed that approximates the $G_{0}W_{0}$ self-energy. This non-local potential defines a new non-interaction Hamiltonian $H_{0}$ that produces a new input Green's function $G_0$. Since the QPsc\textit{GW} concept also requires the calculation of the full non-local $G_{0}W_{0}$ self-energy our expectation is that it will be easier to go beyond $GW$ in our RDMFE framework.

\section{\label{sec:summary}Conclusions}
We have presented an embedding scheme for periodic systems based on Green's functions in the DMFT
framework, which maps an infinite periodic system to a single-site (single unit cell) problem
coupled to an electronic bath that needs to be determined self-consistently. Our RDMFE Green's function 
mapping allows a natural definition of the embedded region
and defines a self-consistency loop which, at convergence, yields self-consistent Green's functions.
The coupling to the surrounding is of dynamical nature enabling electron exchange between the embedded region and the surrounding.
We showed that our scheme produces densities of states and total energies that converge well with increasing size of the
embedded region. We also demonstrated, that the main features of the ``better" theory are rapidly captured 
within our scheme; for example the plasmon satellite already appears in RDMFE@$GW$ calculations for 2 atoms in the Si unit cell.
RDMFE is therefore a promising embedding scheme, that has the potential to make sophisticated and computationally expensive first-principles theories available for periodic systems.

%The actual approximation introduced in our scheme is that the self-energy correction of the higher-level
%method to the effective potential of the lower-level method is assumed to be local. This approximation is 
%a controlled one in the sense that it can be systematically improved by increasing the size of the embedded cluster.
%We consider that our RDMFE will be most useful to systems where the 
%locality approximation to the lattice self-energy is more justified, such as systems with $d$ or $f$-electrons, 
%where KS-DFA is known to give poor descriptions. This potential application will be explored in future work. 
%Moreover, taking the change of the Hartree potential into account by feeding the local self-energy
%obtained from the RDMFE into the a new KS-DFA calculation leading to a new KS Hamiltonian, is an interesting and 
%tractable future goal.

%%%%%%%%%%%%%%%%%%%%%%%%%%%%%%
%%  Beginn der Bibliographie  %%
%%%%%%%%%%%%%%%%%%%%%%%%%%%%%%
% \section*{References}
% \bibliographystyle{unsrt}
% \bibliographystyle{ynt}
\bibliography{./CommonBib.bib}

\begin{thebibliography}{92}
\bibitem [{unf()}]{unfolding}%
  \BibitemOpen
  \href@noop {} {}\bibinfo {note} {Unit cells larger than the primitive unit
  cell give rise to a folded band structure. To cast the band structure back
  onto the Brillouin zone of the 2 atom unit cell we use the unfolding approach
  of L. Nemec.}\BibitemShut {Stop}%
\end{thebibliography}
\end{document}